\documentclass{IEEEtaes}

\usepackage{color,array,amsthm}
\usepackage{graphicx}
\usepackage{bm}
\usepackage{amsmath}
\usepackage{makecell}
\usepackage{algorithm}
\usepackage{algorithmic}
\usepackage{booktabs}
\usepackage{comment}
\usepackage{yhmath}
\jvol{XX}
\jnum{XX}
\jmonth{XXXXX}
\paper{1234567}
\pubyear{2020}
\doiinfo{TAES.2020.Doi Number}

\begin{document}

\title{Joint optimization for DPE/INS deep integration in GNSS
correlation domain}

\author{WEI GAO}
\author{RONG YANG}
\member{Senior Member, IEEE}
\author{JIHONG HUANG}
\author{XINGQUN ZHAN}
\member{Senior Member, IEEE}
\affil{Shanghai Jiao Tong University, Shanghai 200240, China}


\receiveddate{Manuscript received XXXXX 00, 0000; revised XXXXX 00, 0000; accepted XXXXX 00, 0000.\\
This research was supported by the National Natural Science Foundation of China (U2570201). The authors appreciate the assistance of GPT-5.6 Sol and Codex (OpenAI) during manuscript preparation. This work has been submitted to the IEEE for possible publication.}

\corresp{{\itshape (Corresponding author: Rong Yang)}. }

\authoraddress{Wei Gao, Rong Yang, Jihong Huang, and Xingqun Zhan are with the School of Aeronautics and Astronautics, Shanghai Jiao Tong University, Shanghai 200240, China. (e-mail: gaowei1515@sjtu.edu.cn; rongyang@sjtu.edu.cn; jihong.huang@sjtu.edu.cn; xqzhan@sjtu.edu.cn;)}


\markboth{GAO ET AL.}{DPE/INS DEEP INTEGRATION}
\maketitle

\begin{abstract} { Compared with conventional loose/tight/deep integrations of Global Navigation Satellite System (GNSS) and inertial navigation systems (INS), replacing conventional GNSS receivers with Direct Position Estimation (DPE) receivers can exploit richer correlation domain information. However, our previous DPE/INS quasi-deep integration work relies on computationally intensive grid searches and still performs fusion in the navigation domain.} To address these issues, this paper proposes a DPE/INS deep integration mode, in which { the navigation state is jointly optimized using GNSS correlation measurements and INS prior information, instead of fusing a standalone DPE navigation solution with the INS prediction. Building on Line-of-Sight (LOS)-sampled DPE, the computational burden is significantly reduced and the INS aiding information is used to guide the design of sampling parameters.} The proposed deep integration system achieves performance  {approaching} the derived theoretical Posterior Cramér-Rao Lower Bound (PCRLB) in simulation scenarios, and outperforms the DPE/INS quasi-deep integration in navigation domain in urban vehicle experiments.
\end{abstract}


\section{INTRODUCTION}

\begin{figure*} [ht]
    \centering
    \includegraphics[width=1\linewidth]{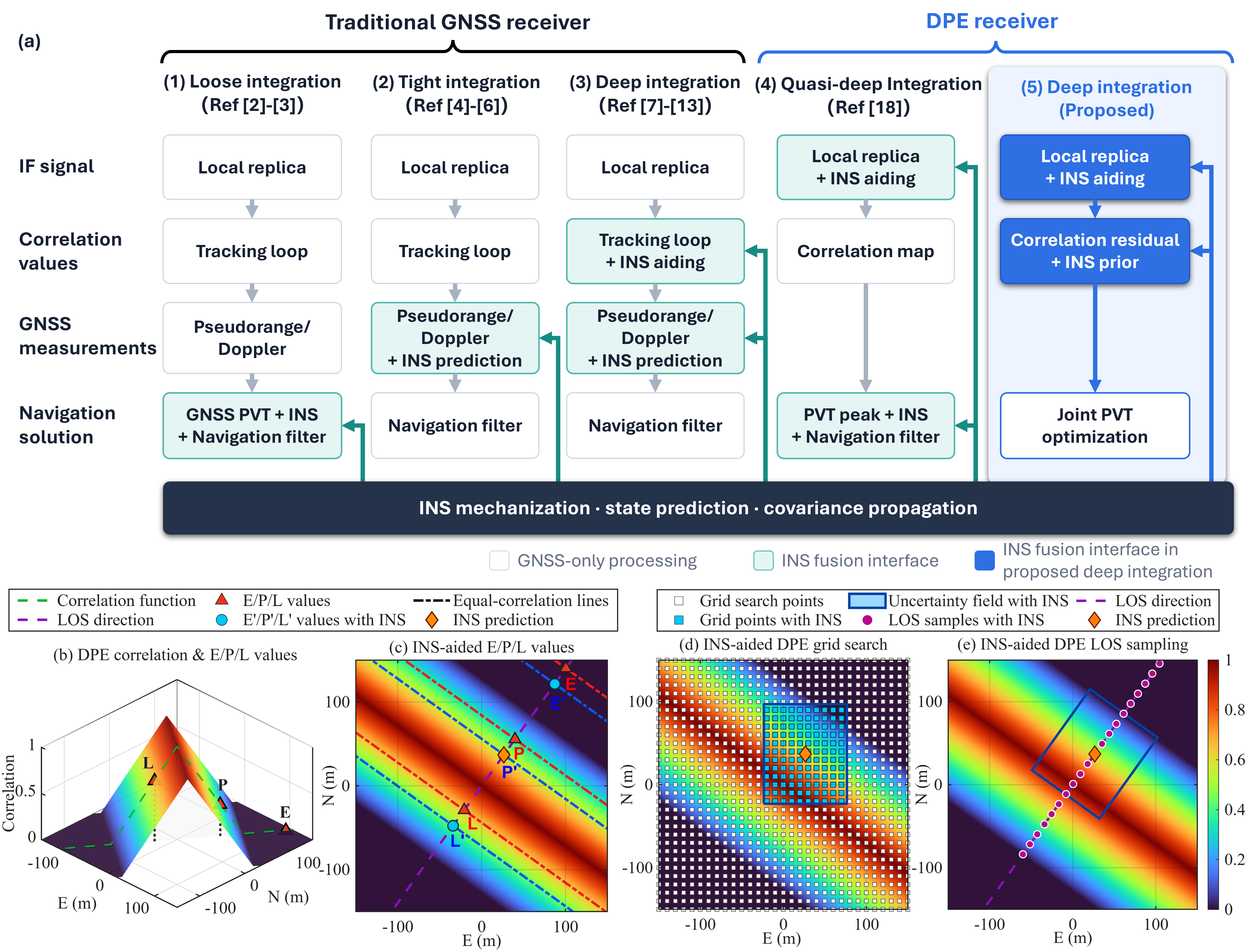}
    \caption{Architecture comparison from conventional GNSS/INS loose/tight/deep integration to DPE/INS quasi-deep/deep integration (a), the correlation domain comparison for conventional GNSS receivers and DPE receivers (b), the INS aiding for conventional GNSS/INS deep integration (c), the INS aiding for DPE/INS quasi-deep integration (d), the INS aiding for DPE/INS deep integration (e)}
    \label{fig:Introduction}
\end{figure*}

The complementary characteristics of Global Navigation Satellite System (GNSS) and inertial navigation systems (INS) establish GNSS/INS integrated navigation as the gold-standard solution in the navigation field \cite{ref:GNSS/INS integration}. The widely applied loose \cite{ref:GNSS/INS software}-\cite{ref:Loose integration network} and tight integration architectures \cite{ref:Tight integration}-\cite{ref:GNSS/INS FGO EKF} operate at the navigation solution level and the GNSS measurement level respectively, as shown in Fig. \ref{fig:Introduction}a(1) and a(2). However, under challenging conditions such as high dynamics, weak signals, and multipath, the degraded solutions and measurements are insufficient to capture these environmental challenges \cite{ref:Fish-eye aiding}. { This motivates integration at an earlier stage of GNSS signal processing.\par}

Correlation values retain more information from the intermediate-frequency  (IF) signal than GNSS measurements \cite{ref:Ultra-tight concept}. Performing GNSS/INS deep integration at the correlation level shown in Fig. \ref{fig:Introduction}a(3) enables the system to extract challenging characteristics that are masked in loose or tight integration modes, thus improving the performance upper bound \cite{ref:Deep_vs_Tight}-\cite{ref:Deep_coupled_urban}. As depicted in Fig. \ref{fig:Introduction}b and c, INS-aided E/P/L correlation values in code tracking can be aligned closer to the correlation peak, thus reducing the probability of anomalies such as signal loss of lock and side-peak locking \cite{ref:Code_tracking_integrity}-\cite{ref:Ultra-tight tracking}. Several studies have quantitatively analyzed the effects of inertial measurement unit (IMU) quality \cite{ref:IMU_quality} and different error sources \cite{ref:Ultra-tight integration} on acquisition and tracking. {These results demonstrate the value of analyzing how INS uncertainty affects correlation-domain processing.\par}

{ Direct position estimation (DPE) offers an alternative receiver architecture that estimates position, velocity, and time (PVT) directly from GNSS correlation values, bypassing pseudorange and Doppler estimation \cite{ref:DPE define}. Its potential benefits have been investigated through theoretical bounds \cite{ref:DPE_CRB} and urban experiments \cite{ref: Urban test}, including methods for multipath mitigation \cite{ref:DPE multipath}.} As shown in Fig. \ref{fig:Introduction}b, whereas the traditional code loop uses only three E/P/L values along the correlation function, the DPE receiver calculates the correlation values across the full PVT space. {Incorporating prior information such as INS model establishes the idea of Bayesian DPE \cite{ref:BDPE original} and can be further implemented by particle filters and square-root derivative-free Gaussian filters \cite{ref:Bayesian DPE IMU}. Subsequent work developed numerically stable particle-weight updates \cite{ref:BDPE particle weight}, nuisance parameters for residual code-delay and Doppler errors \cite{ref:BDPE nuisance}, and efficient probability-map generation for Bayesian DPE \cite{ref:BDPE probability}. Recent urban experiments further demonstrated particle-filter-based Bayesian DPE using a GNSS-only motion model \cite{ref:BDPE urban}. These studies established the Bayesian DPE framework that integrates DPE with other prior information \cite{ref:DPE review}, providing theoretical Bayesian basis for in-depth investigation of how DPE and INS can be fused.\par}

{ From the perspective of Bayesian DPE, our previous work established a DPE/INS quasi-deep integration framework in the navigation domain \cite{ref:Quasi-Deep}, as shown in Fig. \ref{fig:Introduction}a(4).} The INS assistance can significantly reduce the grid-search range of the DPE receiver, focusing the correlation values around the peak region, as shown in Fig. \ref{fig:Introduction}d. {The study analyzed the fusion weights for navigation estimates and search-space reduction in correlation domain under the INS assistance. However, the potential embedded in GNSS correlation values has not yet been fully exploited, since the fusion of DPE and INS is still performed in the navigation domain rather than in the correlation domain. Moreover, the retained grid search scheme maintains a substantial computational burden. Consequently, this framework can only be regarded as quasi-deep integration rather than deep integration. \par}

{To better exploit the PVT information contained in GNSS correlation values, our previous work developed a spatial sparse sampling-based iterative optimization framework for DPE \cite{ref:SS DPE}. This framework treats correlation values as measurements directly related to PVT and estimates PVT through gradient-based minimization of weighted measurement residuals, thereby reducing the reliance on dense grid searches and their associated computational burden \cite{ref:Parallelized_DPE}. Building on this formulation, we developed a per-satellite line-of-sight (LOS) sampling strategy for correlation values , which retains informative correlation measurements while reducing the number of correlation evaluations from exponential growth to linear growth \cite{ref:LOS sampling}. The above redesigns to the estimation criterion and sampling strategy of correlation values lay the theoretical foundation for DPE/INS correlation domain joint optimization framework developed in this paper. \par}

{
Building on these foundations, this paper develops the correlation domain DPE/INS deep integration framework using joint optimization. As illustrated in Fig. \ref{fig:Introduction}a(5), this framework jointly estimates the PVT state using DPE correlation measurements and INS prior information in the correlation domain. The DPE receiver itself no longer outputs a PVT estimate, while the PVT fusion estimation is performed directly in the correlation domain can more fully exploit the potential of GNSS correlation values for the integrated system. On the other hand, as shown in Fig. \ref{fig:Introduction}e, INS prior information can also assist the LOS sampling strategy of correlation values for DPE receivers, reducing the number of samples falling into noise regions and guiding the sampling region to cover the correlation main peak with a specified probability. Direct information fusion at the correlation-domain level, together with efficient and robust sampling and utilization of correlation values, are features that distinguish the proposed framework from DPE/INS quasi-deep integration. \par}

{ The contributions of this paper are summarized as follows. 1) 
The previous DPE/INS quasi-deep integration framework is extended to a correlation-domain DPE/INS deep integration framework based on joint optimization. The theoretical basis includes Bayesian DPE perspective established in \cite{ref:BDPE original}-\cite{ref:DPE review} and the correlation measurements defined in \cite{ref:SS DPE}. 2) Using the Posterior Cramér-Rao Lower Bound (PCRLB) recursion in \cite{ref:PCRLB} and the Fisher information loss analysis for LOS-sampled DPE in \cite{ref:LOS sampling}, the INS prior information and the information provided by the correlation measurements are characterized to establish the relationship between theoretical performance of DPE/INS deep integration and LOS sampling parameters. 3) An INS-aided LOS sampling strategy is designed for DPE/INS deep integration, considering both the fusion information requirement and the correlation peak coverage.} The simulation demonstrates that the theoretical PCRLB of the proposed deep integration framework is consistent with the Monte Carlo experimental results, and also provides the theoretical LOS sampling parameter under INS aiding. Urban vehicle experiments verify that the proposed deep integration framework achieves performance improvements over the quasi-deep integration in challenging scenarios.

The rest of this paper is organized as follows. Section II establishes the proposed DPE/INS deep integration framework in correlation domain. Section III analyzes the theoretical performance for DPE/INS deep integration and develops the INS-aided LOS sampling strategy. Section IV validates the proposed integration framework through simulations and real-world experiment. Finally, Section V concludes this paper.

\section{THE DPE/INS DEEP INTEGRATION FRAMEWORK}

\subsection{The DPE and INS models in correlation domain}

The proposed DPE/INS integration framework is based on a well-defined state-space model. The full system state is captured by the vector ${\bf{x}}$, which includes the key estimation parameters of the INS system and the DPE receiver
\begin{equation}
{\bf{x}} = {\left[ {\begin{array}{*{20}{c}}
{{{\bf{q}}^T}}&{{{\bf{v}}^T}}&{{{\bf{p}}^T}}&{{\bf{b}}_a^T}&{{\bf{b}}_g^T}&{c\delta t}&{c\delta \dot t}
\end{array}} \right]^T}
\label{eq:state_all}
\end{equation}
where ${\bf{q}}$ is the attitude quaternion, ${\bf{v}}$ and ${\bf{p}}$ are the velocity and position vectors. The ${\bf{b}}_a$ and ${\bf{b}}_g$ are accelerometer and gyroscope bias terms, respectively. The receiver clock bias and drift ${\delta t}$ and ${\delta \dot t}$ are normalized by light speed $c$. The time evolution of the state ${\bf{x}}$ is governed by the INS mechanization function and clock model ${\bf{f}}\left(  \cdot  \right)$. The state ${{\bf{x}}_{k}}$ at time epoch $k$ is a function of the previous state ${{\bf{x}}_{k-1}}$ and the IMU measurements ${\bf u}_{k-1}$, perturbed by process noise
\begin{equation}
{{\bf{x}}_k} = {\bf{f}}\left( {{{\bf{x}}_{k - 1}},{{\bf{u}}_{k - 1}}} \right) + {{\bf{w}}_{k - 1}}
\label{eq:x_model}
\end{equation}
The DPE receiver maintains the PVT state of the system \cite{ref:DPE define}, shown as a part of the full state ${\bf{x}}$
\begin{equation}
{\bm{\gamma }} = {\left[ {\begin{array}{*{20}{c}}
{{{\bf{p}}^T}}&{c\delta t}&{{{\bf{v}}^T}}&{c\delta \dot t}
\end{array}} \right]^T} = {{\bf S_{\bm \gamma}}{\bf x}}
\label{eq:DPE_state}
\end{equation}
Then the remaining states in the full state ${\bf{x}}$ that are not directly related to DPE receiver is
\begin{equation}
{\bm{\eta }} = {\left[ {\begin{array}{*{20}{c}}
{{{\bf{q}}^T}}&{{\bf{b}}_a^T}&{{\bf{b}}_g^T}
\end{array}} \right]^T} = {{\bf S_{\bm \eta}}{\bf x}}
\label{eq:remaining_state}
\end{equation}
where ${{\bf S_{\bm \gamma}}}$ and ${\bf S_{\bm \eta}}$ denote the constant selection matrices for sub-states ${\bm{\gamma }}$ and ${\bm{\eta }}$. The snapshot observation of PVT state $\bm \gamma$ at time epoch $k$ is described by the DPE correlation measurement ${{\bf{z}}_k}$
\begin{equation}
{{\bf{z}}_k} = {\bf{h}}\left( {{{\bm{\gamma }}_k}} \right) + {{\bm{\nu }}_k}
\label{eq:z_model}
\end{equation}
where ${\bf{h}}\left(  \cdot  \right)$ is the nonlinear function mapping the DPE state ${{{\bm{\gamma}}_k}}$ to the measurement vector ${{\bf{z}}_k}$, and ${{\bm{\nu }}_k}$ is the measurement noise. Next, we will specifically analyze the state evolution function ${\bf{f}}\left(  \cdot  \right)$ and the process noise $\bf w$ maintained by the INS and clock model, as well as the measurement function 
${\bf{h}}\left(  \cdot  \right)$ and measurement noise ${{\bm{\nu }}}$ maintained by the DPE correlation measurements.

The INS mechanization function is shown as \cite{ref:INS computation}
\begin{equation}
{{\bf{q}}_{k + 1}} = {{\bf{q}}_k} + \frac{1}{2}\int_{{t_k}}^{{t_{k + 1}}} {\left( {{\bf{q}} \circ {\bm{\omega }}_{ib}^{b'} - {\bm{\omega }}_{ie}^{e'} \circ {\bf{q}}} \right)dt} 
\label{eq:q_INS}
\end{equation}
\begin{equation}
{{\bf{v}}_{k + 1}} = {{\bf{v}}_k} + \int_{{t_k}}^{{t_{k + 1}}} {\left( {{\bf{C}}_b^e{{\bf{f}}^b} - 2{\bm{\omega }}_{ie}^e \times {\bf{v}} + {{\bf{g}}^e}} \right)dt}
\label{eq:v_INS}
\end{equation}
\begin{equation}
{{\bf{p}}_{k + 1}} = {{\bf{p}}_k} + \int_{{t_k}}^{{t_{k + 1}}} {{\bf{v}}dt}
\label{eq:p_INS}
\end{equation}
Here ‘b’ denotes the body coordinate and ‘e’ denotes the Earth-centered Earth-fixed (ECEF) coordinate. The IMU measurements ${\bf u}_{k-1}$ in (\ref{eq:x_model}) include the angular rate ${{\bm{\omega }}_{ib}^{b}}$ measured by gyroscopes and specific force ${{{\bf{f}}^b}}$ measured by accelerometers. The Earth rotation rate is ${\bm{\omega }}_{ie}^e = {\left[ {\begin{array}{*{20}{c}}{\rm{0}}&{\rm{0}}&{{\omega _{ie}}}\end{array}} \right]^{\rm{T}}}$ with ${\omega _{ie}} = 7.292 \times {10^{ - 5}} \text{rad/s}$. The rotation matrix ${{\bf{C}}_b^e}$ transforms the b-frame to the e-frame. The superscript ‘$'$’ converts a 3-D vector to the related quaternion, while ‘$ \circ $’ denotes the quaternion multiplication. The receiver clock bias and drift are modeled as
\begin{equation}
c\delta {t_{k + 1}} = c\delta {t_k} + \int_{{t_k}}^{{t_{k + 1}}} {c\delta {{\dot t}_k}dt}
\label{eq:clock_bias_model}
\end{equation}
\begin{equation}
c\delta {\dot t_{k + 1}} = c\delta {\dot t_k} - \frac{1}{{{T_{clk}}}}\int_{{t_k}}^{{t_{k + 1}}} {c\delta {{\dot t}_k}dt} 
\label{eq:clock_drift_model}
\end{equation}
The factor ${{T_{clk}}}$ is the time constant of first-order Gauss-Markov (GM) process for clock drift $c\delta {\dot t}$. The equations (\ref{eq:q_INS})-(\ref{eq:clock_drift_model}) together constitute the state evolution function ${\bf{f}}\left(  \cdot  \right)$ in (\ref{eq:x_model}). The process noise is then given as \cite{ref:Quasi-Deep}
\begin{equation}
{\bf{w}} = {\left[ {\begin{array}{*{20}{c}}
{\frac{1}{2}{{\left( {{\bf{U}}{{\bf{n}}_{rg}}} \right)}^T}}&{{{\left( {{\bf{C}}_b^e{{\bf{n}}_{ra}}} \right)}^T}}&{\bf{0}}&{{\bf{n}}_a^T}&{{\bf{n}}_g^T}&{n_b^T}&{n_d^T}
\end{array}} \right]^T}
\label{eq:noise_model}
\end{equation}
where $\bf U$ is linearized coefficient for attitude quaternion $\bf q$, while ${{\bf{n}}_{ra}}$ and ${{\bf{n}}_{rg}}$ represent the Gaussian white noises of gyroscopes output ${{\bm{\omega }}_{ib}^{b}}$ and accelerometers output ${{{\bf{f}}^b}}$, respectively. Similarly, the Gaussian white noise ${{\bf{n}}_a}$, ${{\bf{n}}_g}$, ${{n}_b}$, and ${{n}_d}$ serve as the driving noise for accelerometers bias ${\bf{b}}_a$, gyroscope bias ${\bf{b}}_g$, clock bias $c\delta t$ and clock drift $c\delta \dot t$. Consequently, the covariance matrix ${{\bf{Q}}_k}$ of the process noise ${{\bf{w}}_k}$ is obtained as ${{\bf{Q}}_k} = E\left( {{{\bf{w}}_k}{\bf{w}}_k^T} \right)$.

The DPE correlation values from all spatial sampling points across each satellite channel constitute the measurement vector ${{\bf{z}}_k}$ in (\ref{eq:z_model}), defined as \cite{ref:SS DPE}
\begin{equation}
{\bf{z}} = {\left[ {{\Re ^1}\left( {{{{\bm{\tilde \gamma }}}^{1,1}}} \right), \cdots ,{\Re ^1}\left( {{{{\bm{\tilde \gamma }}}^{1,L}}} \right), \cdots ,{\Re ^M}\left( {{{{\bm{\tilde \gamma }}}^{M,L}}} \right)} \right]^T}
\label{eq:z_define}
\end{equation}
where the superscript “$\sim$” denotes the spatial sampling points of the DPE, and the $l$-th spatial sampling point of the $m$-th satellite is expressed as ${{{{\bm{\tilde \gamma }}}^{m,l}}}$.  The complete set of spatial sampling points $\left\{ {{{{\bm{\tilde \gamma }}}^{1,1}}, \cdots {{{\bm{\tilde \gamma }}}^{1,L}}, \cdots ,{{{\bm{\tilde \gamma }}}^{M,L}}} \right\}$ can be obtained through the LOS sampling strategy. Then the correlation value corresponding to ${{{{\bm{\tilde \gamma }}}^{m,l}}}$ is obtained as
\begin{equation}
{\Re ^m}\left( {{{{\bm{\tilde \gamma }}}^{m,l}}} \right) = \frac{1}{N}\sum\limits_{n = 1}^N {s\left[ n \right]{{\tilde s}^m}\left[ {n,{{{\bm{\tilde \gamma }}}^{m,l}}} \right]}
\label{eq:Correlation_define}
\end{equation}
where ${s\left[ n \right]}$ denotes the raw GNSS IF signal, and $N$ is the length of the signal data within the coherent integration time $T_c$, satisfying $N = {f_s}{T_c}$ with $f_s$ being the signal sampling frequency. The term ${{{\tilde s}^m}\left[ {n,{{{\bm{\tilde \gamma }}}^{m,l}}} \right]}$ represents the locally replicated signal corresponding to ${{{{\bm{\tilde \gamma }}}^{m,l}}}$. 

The correlation value can be modeled as an analytical part and a noise part
\begin{equation}
{\Re ^m}\left( {{{{\bm{\tilde \gamma }}}^{m,l}}} \right) = {h^m}\left( {{{{\bm{\tilde \gamma }}}^{m,l}},{\bm{\gamma }}} \right) + {\nu ^{m,l}}
\label{eq:Correlation_model}
\end{equation}
The analytical part in (\ref{eq:Correlation_model}) is derived as \cite{ref:SS DPE}
\begin{equation}
{h^m}\left( {{{{\bm{\tilde \gamma }}}^{m,l}},{\bm{\gamma }}} \right) = \frac{1}{2}{A^m}\Re _\tau ^m\left( {\Delta {\tau ^{m,l}}} \right)\Re _f^m\left( {\Delta f^{m,l}} \right)\exp \left( {j{\varphi ^m}} \right)
\label{eq:h_model}
\end{equation}
where ${A^m}$ is the signal amplitude, and ${\varphi^m}$ is the carrier phase. The error terms $\Delta {\tau ^{m,l}} = {\tau ^m} - {\tilde \tau ^{m,l}}$ and $\Delta {f^{m,l}} = {f^m} - {\tilde f^{m,l}}$ represent the code-delay and Doppler errors between sample points and the truth. The correlation functions $\Re _\tau ^m\left( \cdot \right)$ and $\Re _f^m\left( \cdot \right)$ are given as
\begin{equation}
\Re _\tau ^m\left( {\Delta {\tau ^m}} \right) \approx \max \left( {0,1 - \left| {\Delta {\tau ^m}} \right|} \right)
\label{eq:tao_function}
\end{equation}
\begin{equation}
\Re _f^m\left( {\Delta {f^m}} \right) = \frac{{{\rm{sin}}\left( {\pi \Delta {f^m}{T_c}} \right)}}{{\pi \Delta {f^m}{T_c}}} \buildrel\textstyle.\over= {\rm{sinc}}\left( {\Delta {f^m}{T_c}} \right)
\label{eq:doppler_function}
\end{equation}
The covariance of noise part in (\ref{eq:Correlation_model}) is shown as
\begin{equation}
E\left[ {{\nu ^{m,i}}{{\left( {{\nu ^{m,j}}} \right)}^*}} \right] \approx \frac{{{N_0}}}{{4{T_c}}}\Re _\tau ^m\left( {\nabla \tau _{i,j}^m} \right)\Re _f^m\left( {\nabla f_{i,j}^m} \right)
\label{eq:noise_covariance}
\end{equation}
The coherent integration time $T_{c}$ and noise power spectral density $N_0$ are constant parameters. The difference terms $\nabla \tau _{i,j}^m = {\tilde \tau ^{m,j}} - {\tilde \tau ^{m,i}}$ and $\nabla f_{i,j}^m = {\tilde f^{m,j}} - {\tilde f^{m,i}}$ represent the differences of code-delay and Doppler between sample points ${{{{\bm{\tilde \gamma }}}^{m,j}}}$ and ${{{{\bm{\tilde \gamma }}}^{m,i}}}$. The code-delay ${\tau ^m}$ and Doppler ${f^m}$ in (\ref{eq:h_model})-(\ref{eq:noise_covariance}) are connected to PVT state $\bm \gamma$ by \cite{ref:Geometric DPE}
\begin{equation}
{\tau ^m} = \frac{{{f_c}}}{c}\left( {\left\| {{{\bf{p}}^m_s} - {\bf{p}}} \right\| + c\delta t - c\delta {t^m_s}} \right)
\label{eq:tao_define}
\end{equation}
\begin{equation}
{f^m} =  - \frac{{{f_L}}}{c}\left[ {\frac{{{{\left( {{\bf{p}}_s^m - {\bf{p}}} \right)}^T}}}{{\left\| {{\bf{p}}_s^m - {\bf{p}}} \right\|}}\left( {{\bf{v}}_s^m - {\bf{v}}} \right) + c\delta \dot t - c\delta \dot t_s^m} \right]
\label{eq:doppler_define}
\end{equation}
where $f_c$ is the PRN code frequency and $f_L$ is the carrier frequency. The subscript ‘$s$’ denotes the satellite-related parameters. The sampled code-delay ${\tilde \tau ^{m,l}}$ and Doppler ${\tilde f^{m,l}}$ can be obtained by substituting sampling point ${{{{\bm{\tilde \gamma }}}^{m,l}}}$ into (\ref{eq:tao_define}) and (\ref{eq:doppler_define}).

The vector ${\bf h}\left(\cdot\right)$ and $\bm \nu$ in (\ref{eq:z_model}) collect the analytical part and noise part in (\ref{eq:Correlation_model}) for all spatial sampling points $\left\{ {{{{\bm{\tilde \gamma }}}^{1,1}}, \cdots {{{\bm{\tilde \gamma }}}^{1,L}}, \cdots ,{{{\bm{\tilde \gamma }}}^{M,L}}} \right\}$. The measurement noise covariance has the block-diag form
\begin{equation}
{{\bf{R}}_k} = E\left( {{{\bm{\nu}}_k}{\bm{\nu}}_k^T} \right) =  diag\left( {{{\left[ {{{\bf{R}}^1},{{\bf{R}}^2}, \cdots ,{{\bf{R}}^M}} \right]}_k}} \right)
\label{eq:R_define}
\end{equation}
For each satellite block ${{{\bf{R}}^m}}$ of the covariance matrix in (\ref{eq:R_define}), the element at the $i$-th row and $j$-th column is given as equation (\ref{eq:noise_covariance}).

In summary, the INS model and the clock model maintain the state propagation process between epochs, characterizing the time evolution of the system, while the DPE model maintains the relationship between the correlation values of spatial sampling points and the system state, characterizing the spatial sampling characteristics of the system.

\subsection{The estimation framework for DPE/INS deep integration}

Before constructing the DPE/INS deep integration framework, we first define the notation for the estimated states, estimation errors, and estimation covariances of the INS and DPE/INS fusion systems, respectively, to facilitate subsequent derivations. Note that within the DPE/INS deep integration framework, the standalone DPE estimates and covariances are not calculated, and therefore no corresponding definitions are provided. These definitions are summarized in Table \ref{tab:Error_definition}.

\begin{table}[ht]
    \centering
        \caption{The notation used in DPE/INS deep integration framework}
    \begin{tabular}{|c|c|c|c|}\hline
         Type&Estimation& Estimation error&Estimation covariance\\\hline
 INS&${{\bf{\mathord{\buildrel{\lower3pt\hbox{$\scriptscriptstyle\frown$}} 
\over x} }}_k}$& $\delta {{\bf{x}}_k} = {{\bf{x}}_k} - {{\bf{\mathord{\buildrel{\lower3pt\hbox{$\scriptscriptstyle\frown$}} 
\over x} }}_k}$&${{\bf{\mathord{\buildrel{\lower3pt\hbox{$\scriptscriptstyle\frown$}} 
\over P} }}_k} = E\left( {\delta {{\bf{x}}_k}\delta {\bf{x}}_k^T} \right)$\\\hline
 Fusion&${{\bf{\hat x}}_k}$& $\Delta {{\bf{x}}_k} = {{\bf{x}}_k} - {{\bf{\hat x}}_k}$&${{\bf{\hat P}}_k} = E\left( {\Delta {{\bf{x}}_k}\Delta {\bf{x}}_k^T} \right)$\\\hline
    \end{tabular}
    \label{tab:Error_definition}
\end{table}

The proposed DPE/INS deep integration framework is formulated within a Bayesian estimation framework. Let ${{\bm{{\rm Z}}}_k}$ denote the collection of all DPE measurement vectors up to epoch $k$, defined as
\begin{equation}
{{\bm{{\rm Z}}}_k} = \left\{ {{{\bf{z}}_1},{{\bf{z}}_2}, \cdots ,{{\bm{{\rm z}}}_k}} \right\}
\label{eq:Z_set}
\end{equation}
The optimal estimate of the state ${{{\bf{x}}_k}}$ is then obtained by maximizing the posterior probability density function $p\left( {{{\bf{x}}_k}|{{\bm{{\rm Z}}}_k}} \right)$. This maximum a posteriori (MAP) estimate is given by
\begin{equation}
{{\bf{\hat x}}_k} = \mathop {\arg \max }\limits_{{{\bf{x}}_k}} p\left( {{{\bf{x}}_k}|{{\bm{{\rm Z}}}_k}} \right)
\label{eq:MAP_define}
\end{equation}
Applying Bayes' theorem and the chain rule of conditional probability, the posterior distribution can be expressed as
\begin{equation}
\begin{aligned}
p\left( {{{\bf{x}}_k}|{{\bm{{\rm Z}}}_k}} \right) &= \frac{{p\left( {{{\bf{z}}_k}|{{\bf{x}}_k},{{\bm{{\rm Z}}}_{k - 1}}} \right)p\left( {{{\bf{x}}_k}|{{\bm{{\rm Z}}}_{k - 1}}} \right)}}{{p\left( {{{\bf{z}}_k}|{{\bm{{\rm Z}}}_{k - 1}}} \right)}}\\
 &\propto p\left( {{{\bf{z}}_k}|{{\bm{\gamma }}_k}} \right)p\left( {{{\bf{x}}_k}|{{\bm{{\rm Z}}}_{k - 1}}} \right)
 \end{aligned}
\label{eq:MAP_prob}
\end{equation}
The second line follows from the simplification $p\left( {{{\bf{z}}_k}|{{\bf{x}}_k},{{\bm{{\rm Z}}}_{k - 1}}} \right) = p\left( {{{\bf{z}}_k}|{{\bm{\gamma }}_k}} \right)$, which indicates the fact that, conditioned on the current PVT state ${{\bm{\gamma}}_k}$, the current measurement ${{\bf z}_k}$ is independent of both the remaining state components ${\bm \eta}_k$ and all earlier measurements ${{\bf Z}_{k-1}}$.

Substituting (\ref{eq:MAP_prob}) into (\ref{eq:MAP_define}), the MAP estimation criterion simplifies to
\begin{equation}
\begin{aligned}
{{{\bf{\hat x}}}_k} 
 =& \mathop {\arg \max }\limits_{{{\bf{x}}_k}} p\left( {{{\bf{z}}_k}|{{\bm{\gamma }}_k}} \right)p\left( {{{\bf{x}}_k}|{{\bm{{\rm Z}}}_{k - 1}}} \right)\\
 =& \mathop {\arg \min }\limits_{{{\bf{x}}_k}} \left( {\left\| {{{\bf{z}}_k} - {\bf{h}}\left( {{{\bm{\gamma }}_k}} \right)} \right\|_{{\bf{R}}_k^{ - 1}}^2 + \left\| {{{\bf{x}}_k} - {{{\bf{\mathord{\buildrel{\lower3pt\hbox{$\scriptscriptstyle\frown$}} 
\over x} }}}_k}} \right\|_{{\bf{\mathord{\buildrel{\lower3pt\hbox{$\scriptscriptstyle\frown$}} 
\over P} }}_k^{ - 1}}^2} \right)
\end{aligned}
\label{eq:MAP_criterion}
\end{equation}
where the last equality in (\ref{eq:MAP_criterion}) follows from the Gaussian assumptions for distribution $p\left( {{{\bf{z}}_k}|{{\bm{\gamma}}_k}} \right)$ and $p\left( {{{\bf{x}}_k}|{{\bm{{\rm Z}}}_{k - 1}}} \right)$. The formulation of (\ref{eq:MAP_criterion}) constitutes the core optimization criterion for the proposed DPE/INS deep integration framework, where the two terms ${\left\| {{{\bf{z}}_k} - {\bf{h}}\left( {{{\bm{\gamma}}_k}} \right)} \right\|_{{\bf{R}}_k^{ - 1}}^2}$ and ${\left\| {{{\bf{x}}_k} - {{{\bf{\mathord{\buildrel{\lower3pt\hbox{$\scriptscriptstyle\frown$}} 
\over x} }}}_k}} \right\|_{{\bf{\mathord{\buildrel{\lower3pt\hbox{$\scriptscriptstyle\frown$}} 
\over P} }}_k^{ - 1}}^2}$ represent contributions from the DPE measurement and the INS prediction, respectively.

Based on the MAP estimation criterion of (\ref{eq:MAP_criterion}), we first analyze the estimation results ${{\bf{\mathord{\buildrel{\lower3pt\hbox{$\scriptscriptstyle\frown$}} 
\over x} }}_k}$ and estimation covariance ${{{\bf{\mathord{\buildrel{\lower3pt\hbox{$\scriptscriptstyle\frown$}} 
\over P} }}}_k}$ provided by INS. Specifically, the predicted state ${{\bf{\mathord{\buildrel{\lower3pt\hbox{$\scriptscriptstyle\frown$}} 
\over x} }}_k}$ at epoch $k$ is obtained by propagating the previous fused state ${{{{\bf{\hat x}}}_{k - 1}}}$ through the INS mechanization
\begin{equation}
{{\bf{\mathord{\buildrel{\lower3pt\hbox{$\scriptscriptstyle\frown$}} 
\over x} }}_k} = {\bf{f}}\left( {{{{\bf{\hat x}}}_{k - 1}},{{\bf{u}}_{k - 1}}} \right)
\label{eq:INS_derive}
\end{equation}
where ${{{{\bf{\hat x}}}_{k - 1}}}$ is the fusion state from the previous epoch $k-1$. Subtracting (\ref{eq:INS_derive}) from (\ref{eq:x_model}) yields the expression for the INS estimation error $\delta {{\bf{x}}_k}$
\begin{equation}
\begin{aligned}
\delta {{\bf{x}}_k} &= {{\bf{x}}_k} - {{{\bf{\mathord{\buildrel{\lower3pt\hbox{$\scriptscriptstyle\frown$}} 
\over x} }}}_k}\\
 &= {\bf{f}}\left( {{{\bf{x}}_{k - 1}},{{\bf{u}}_{k - 1}}} \right) + {{\bf{w}}_{k - 1}} - {\bf{f}}\left( {{{{\bf{\hat x}}}_{k - 1}},{{\bf{u}}_{k - 1}}} \right)\\
&={{{\bf{\hat F}}}_{k - 1}}\Delta {{\bf{x}}_{k - 1}} + {{\bf{w}}_{k - 1}}
\end{aligned}
\label{eq:INS_error}
\end{equation}
where $\Delta {{\bf{x}}_{k - 1}}$ is fusion error at epoch $k-1$, and ${{{\bf{\hat F}}}_{k - 1}}$ is the Jacobian of the state transition function ${\bf{f}}\left(  \cdot  \right)$ at the fusion state ${{{{\bf{\hat x}}}_{k - 1}}}$. This linearized error propagation forms the basis for computing the prediction covariance. The covariance matrix of the predicted state is then given by 
\begin{equation}
{{{\bf{\mathord{\buildrel{\lower3pt\hbox{$\scriptscriptstyle\frown$}} 
\over P} }}}_k} = E\left( {\delta {{\bf{x}}_k}\delta {\bf{x}}_k^T} \right)= {{\bf{\hat F}}_{k - 1}}{{{\bf{\hat P}}}_{k - 1}}{\bf{\hat F}}_{k - 1}^T + {{\bf{Q}}_{k - 1}}
\label{eq:INS_covariance}
\end{equation}
where ${{{\bf{\hat P}}}_{k - 1}}$ is the fusion covariance from the previous epoch.

The INS prediction covariance matrix in \eqref{eq:INS_covariance} is partitioned into blocks conforming to the PVT state $\boldsymbol{\gamma}$ and remaining state $\bm \eta$
\begin{equation}
{{\bf{\mathord{\buildrel{\lower3pt\hbox{$\scriptscriptstyle\frown$}} \over P} }}_k}  \buildrel \Delta \over = {\left[ {\begin{array}{*{20}{c}}
{{{\bf{\mathord{\buildrel{\lower3pt\hbox{$\scriptscriptstyle\frown$}} \over P} }}_{{\bm{\eta \eta }}}}}&{{{\bf{\mathord{\buildrel{\lower3pt\hbox{$\scriptscriptstyle\frown$}} \over P} }}_{{\bm{\eta \gamma }}}}}\\
{{{\bf{\mathord{\buildrel{\lower3pt\hbox{$\scriptscriptstyle\frown$}} \over P} }}_{{\bm{\gamma \eta }}}}}&{{{\bf{\mathord{\buildrel{\lower3pt\hbox{$\scriptscriptstyle\frown$}} \over P} }}_{{\bm{\gamma \gamma }}}}}
\end{array}} \right]_k}
\label{eq:P_block}
\end{equation}
The block-partitioned form of ${{\bf{\mathord{\buildrel{\lower3pt\hbox{$\scriptscriptstyle\frown$}} \over P} }}_k}$ in (\ref{eq:P_block}) is conceptual and assumes a reordered state vector for derivation convenience. The corresponding sub-blocks are defined as ${{\bf{\mathord{\buildrel{\lower3pt\hbox{$\scriptscriptstyle\frown$}} 
\over P} }}_{{\bm{\eta \eta }}}} = {{\bf{S}}_{\bm{\eta }}}{{\bf{\mathord{\buildrel{\lower3pt\hbox{$\scriptscriptstyle\frown$}} 
\over P} }}_k}{\bf{S}}_{\bm{\eta }}^T$, ${{\bf{\mathord{\buildrel{\lower3pt\hbox{$\scriptscriptstyle\frown$}} 
\over P} }}_{{\bm{\eta \gamma }}}} = {{\bf{S}}_{\bm{\eta }}}{{\bf{\mathord{\buildrel{\lower3pt\hbox{$\scriptscriptstyle\frown$}} 
\over P} }}_k}{\bf{S}}_{\bm{\gamma }}^T$, ${{\bf{\mathord{\buildrel{\lower3pt\hbox{$\scriptscriptstyle\frown$}} 
\over P} }}_{{\bm{\gamma \eta }}}} = {{\bf{S}}_{\bm{\gamma }}}{{\bf{\mathord{\buildrel{\lower3pt\hbox{$\scriptscriptstyle\frown$}} 
\over P} }}_k}{\bf{S}}_{\bm{\eta }}^T$, and ${{\bf{\mathord{\buildrel{\lower3pt\hbox{$\scriptscriptstyle\frown$}} 
\over P} }}_{{\bm{\gamma \gamma }}}} = {{\bf{S}}_{\bm{\gamma }}}{{\bf{\mathord{\buildrel{\lower3pt\hbox{$\scriptscriptstyle\frown$}} 
\over P} }}_k}{\bf{S}}_{\bm{\gamma }}^T$. Applying the Schur complement \cite{ref:Schur} to the quadratic form of the INS prior distribution, the item ${\left\| {{{\bf{x}}_k} - {{{\bf{\mathord{\buildrel{\lower3pt\hbox{$\scriptscriptstyle\frown$}} 
\over x} }}}_k}} \right\|_{{\bf{\mathord{\buildrel{\lower3pt\hbox{$\scriptscriptstyle\frown$}} 
\over P} }}_k^{ - 1}}^2}$ in \eqref{eq:MAP_criterion} is equivalently decomposed as
\begin{equation}
\begin{aligned}
&\left\| {{{\bf{x}}_k} - {{\bf{\mathord{\buildrel{\lower3pt\hbox{$\scriptscriptstyle\frown$}} \over x} }}_k}} \right\|_{{\bf{\mathord{\buildrel{\lower3pt\hbox{$\scriptscriptstyle\frown$}} \over P} }}_k^{ - 1}}^2 = \left\| {{{\bm{\gamma }}_k} - {{\bm{\mathord{\buildrel{\lower3pt\hbox{$\scriptscriptstyle\frown$}} \over \gamma } }}_k}} \right\|_{{\bf{\mathord{\buildrel{\lower3pt\hbox{$\scriptscriptstyle\frown$}} \over P} }}_{{\bm{\gamma \gamma }}}^{ - 1}}^2 \\
&+ \left\| {{{\bm{\eta }}_k} - {{\bm{\mathord{\buildrel{\lower3pt\hbox{$\scriptscriptstyle\frown$}} \over \eta } }}_k} - {{\bf{\mathord{\buildrel{\lower3pt\hbox{$\scriptscriptstyle\frown$}} \over P} }}_{{\bm{\eta \gamma }}}}{\bf{\mathord{\buildrel{\lower3pt\hbox{$\scriptscriptstyle\frown$}} \over P} }}_{{\bm{\gamma \gamma }}}^{ - 1}\left( {{{\bm{\gamma }}_k} - {{\bm{\mathord{\buildrel{\lower3pt\hbox{$\scriptscriptstyle\frown$}} \over \gamma } }}_k}} \right)} \right\|_{{\bf{\mathord{\buildrel{\lower3pt\hbox{$\scriptscriptstyle\frown$}} \over P} }}_{{\bm{\eta }}|{\bm{\gamma }}}^{ - 1}}^2
\end{aligned}
\label{eq:State_Schur}
\end{equation}
where the conditional covariance ${{\bf{\mathord{\buildrel{\lower3pt\hbox{$\scriptscriptstyle\frown$}} \over P} }}_{{\bm{\eta }}|{\bm{\gamma }}}}$ is defined as
\begin{equation}
{{\bf{\mathord{\buildrel{\lower3pt\hbox{$\scriptscriptstyle\frown$}} \over P} }}_{{\bm{\eta }}|{\bm{\gamma }}}} = {{\bf{\mathord{\buildrel{\lower3pt\hbox{$\scriptscriptstyle\frown$}} \over P} }}_{{\bm{\eta \eta }}}} - {{\bf{\mathord{\buildrel{\lower3pt\hbox{$\scriptscriptstyle\frown$}} \over P} }}_{{\bm{\eta \gamma }}}}{\bf{\mathord{\buildrel{\lower3pt\hbox{$\scriptscriptstyle\frown$}} \over P} }}_{{\bm{\gamma \gamma }}}^{ - 1}{{\bf{\mathord{\buildrel{\lower3pt\hbox{$\scriptscriptstyle\frown$}} \over P} }}_{{\bm{\gamma \eta }}}}
\label{eq:Condition_cov}
\end{equation}
Equation \eqref{eq:State_Schur} separates the INS prior term into a marginal term involving only ${{\bm\gamma}_k}$ and a nonnegative conditional term coupling ${{\bm\eta}_k}$ to ${{\bm\gamma}_k}$. Since the DPE likelihood also depends only on ${{\bm\gamma}_k}$, the conditional term is minimized, for every fixed ${{\bm\gamma}_k}$, at its conditional mean and therefore becomes zero. The full-state MAP problem consequently reduces exactly to the following nonlinear optimization problem
\begin{equation}
{{\bm{\hat \gamma }}_k} = \mathop {\arg \min }\limits_{{{\bm{\gamma }}_k}} \left( {\left\| {{{\bf{z}}_k} - {\bf{h}}\left( {{{\bm{\gamma }}_k}} \right)} \right\|_{{\bf{R}}_k^{ - 1}}^2 + \left\| {{{\bm{\gamma }}_k} - {{\bm{\mathord{\buildrel{\lower3pt\hbox{$\scriptscriptstyle\frown$}} \over \gamma } }}_k}} \right\|_{{\bf{\mathord{\buildrel{\lower3pt\hbox{$\scriptscriptstyle\frown$}} \over P} }}_{{\bm{\gamma \gamma }}}^{ - 1}}^2} \right)
\label{eq:MAP_PVT}
\end{equation}
Equation \eqref{eq:MAP_PVT} is the core optimization criterion of the proposed framework. Only the DPE-observable sub-state ${{\bm\gamma}_k}$ is optimized numerically, and the remaining INS state ${\bm \eta}_k$ is recovered afterward through their prior cross-covariance with ${{\bm\gamma}_k}$.

To solve the nonlinear optimization problem in \eqref{eq:MAP_PVT}, gradient-based iterative optimization methods can be employed, such as Gauss-Newton method or Levenberg–Marquardt (L-M) method. When we obtained the optimal point ${{{\bm{\hat \gamma }}}_k}$ by these optimization methods, the gradient of the cost function with respect to $\boldsymbol{\gamma}_k$ must vanish, yielding the first-order optimality condition
\begin{equation}
\begin{aligned}
&\partial \left( {\left\| {{{\bf{z}}_k} - {\bf{h}}\left( {{{\bm{\gamma }}_k}} \right)} \right\|_{{\bf{R}}_k^{ - 1}}^2 + \left\| {{{\bm{\gamma }}_k} - {{\bm{\mathord{\buildrel{\lower3pt\hbox{$\scriptscriptstyle\frown$}} \over \gamma } }}_k}} \right\|_{{\bf{\mathord{\buildrel{\lower3pt\hbox{$\scriptscriptstyle\frown$}} \over P} }}_{{\bm{\gamma \gamma }}}^{ - 1}}^2} \right)/\partial {{\bm{\hat \gamma }}_k}\\
=& - 2{\bf{J}}_k^T{\bf{R}}_k^{ - 1}\left[ {{{\bf{z}}_k} - {\bf{h}}\left( {{{\bm{\hat \gamma }}_k}} \right)} \right] + 2{\bf{\mathord{\buildrel{\lower3pt\hbox{$\scriptscriptstyle\frown$}} \over P} }}_{{\bm{\gamma \gamma }}}^{ - 1}\left( {{{\bm{\hat \gamma }}_k} - {{\bm{\mathord{\buildrel{\lower3pt\hbox{$\scriptscriptstyle\frown$}} \over \gamma } }}_k}} \right) = {\bf{0}}
\end{aligned}
\label{eq:Opt_condition}
\end{equation}
where ${\bf{J}}_k = \left. {\partial {\bf{h}}/\partial {\boldsymbol{\gamma}}} \right|_{{\boldsymbol{\gamma}} = {\boldsymbol{\hat \gamma}}_k}$ is the Jacobian matrix of the measurement function evaluated at the optimal point. Linearizing the measurement function ${\bf{h}}\left( {{{\boldsymbol{\gamma }}_k}} \right)$ around the optimal point ${{{\bm{\hat \gamma }}}_k}$ as ${\bf{h}}\left( {{{\boldsymbol{\hat \gamma }}_k}} \right) \approx {\bf{h}}\left( {{{\boldsymbol{\gamma }}_k}} \right) + {{\bf{J}}_k}\Delta {{\boldsymbol{\gamma }}_k}$, with $\Delta {{\boldsymbol{\gamma }}_k} = {\boldsymbol{ \gamma}}_k - {\boldsymbol{\hat \gamma}}_k$ and $\delta {{\boldsymbol{\gamma }}_k} = {\boldsymbol{ \gamma}}_k - {\boldsymbol{\mathord{\buildrel{\lower3pt\hbox{$\scriptscriptstyle\frown$}} \over \gamma } }}_k$, and substituting into \eqref{eq:Opt_condition}, we obtain the linearized error-state relationship expressed as
\begin{equation}
\left( {{\bf{J}}_k^T{\bf{R}}_k^{ - 1}{{\bf{J}}_k} + {\bf{\mathord{\buildrel{\lower3pt\hbox{$\scriptscriptstyle\frown$}} \over P} }}_{{\bm{\gamma \gamma }}}^{ - 1}} \right)\Delta {{\bm{\gamma }}_k} = {\bf{J}}_k^T{\bf{R}}_k^{ - 1}{{\bm{\nu }}_k} + {\bf{\mathord{\buildrel{\lower3pt\hbox{$\scriptscriptstyle\frown$}} \over P} }}_{{\bm{\gamma \gamma }}}^{ - 1}\delta {{\bm{\gamma }}_k}
\label{eq:error_relationship}
\end{equation}
where ${{\bm{\nu }}_k} = {\bf{z}}_k - {\bf{h}}({\boldsymbol{\gamma}}_k)$ is the measurement noise. Under the assumption that the prior error $\delta {{\boldsymbol{\gamma }}_k}$ and the measurement noise ${{\boldsymbol{\nu }}_k}$ are uncorrelated, the posterior covariance of the optimal PVT estimation ${{{\bm{\hat \gamma }}}_k}$ can be obtained from  \eqref{eq:error_relationship}
\begin{equation}
{{\bf{\hat P}}_{{\bm{\gamma \gamma }}}} = E\left( {\Delta {{\bm{\gamma }}_k}\Delta {\bm{\gamma }}_k^T} \right) = {\left( {{\bf{J}}_k^T{\bf{R}}_k^{ - 1}{{\bf{J}}_k} + {\bf{\mathord{\buildrel{\lower3pt\hbox{$\scriptscriptstyle\frown$}} \over P} }}_{{\bm{\gamma \gamma }}}^{ - 1}} \right)^{ - 1}}
\label{eq:Fusion_PVT_cov}
\end{equation}
Equation \eqref{eq:Fusion_PVT_cov} is the information-filter form of the covariance update, where the prior information matrix ${\bf{\mathord{\buildrel{\lower3pt\hbox{$\scriptscriptstyle\frown$}} \over P} }}_{{\bm{\gamma \gamma }}}^{ - 1}$ is augmented by the measurement information ${\bf{J}}_k^T{\bf{R}}_k^{ - 1}{{\bf{J}}_k}$. The covariance ${{\bf{\hat P}}_{{\bm{\gamma \gamma }}}}$ is a local approximation evaluated at the converged solution ${{{\bm{\hat \gamma }}}_k}$ for the optimization problem (\ref{eq:MAP_PVT}).

Once the reduced-dimension estimate ${{\bm{\hat \gamma }}_k}$ and its covariance ${{\bf{\hat P}}_{{\bm{\gamma \gamma }}}}$ are obtained, we can recover the full state estimation by leveraging the conditional Gaussian property implied by the Schur complement decomposition. Specifically, when the optimal PVT state ${{\bm{\hat \gamma }}_k}$ is obtained, the optimal ${\boldsymbol{\hat \eta}}_k$ is given by the conditional mean
\begin{equation}
{{\bm{\hat \eta }}_k} = {\bm{\mathord{\buildrel{\lower3pt\hbox{$\scriptscriptstyle\frown$}} \over \eta } }_k} + {{\bf{\mathord{\buildrel{\lower3pt\hbox{$\scriptscriptstyle\frown$}} \over P} }}_{{\bm{\eta \gamma }}}}{\bf{\mathord{\buildrel{\lower3pt\hbox{$\scriptscriptstyle\frown$}} \over P} }}_{{\bm{\gamma \gamma }}}^{ - 1}\left( {{{\bm{\hat \gamma }}_k} - {{\bm{\mathord{\buildrel{\lower3pt\hbox{$\scriptscriptstyle\frown$}} \over \gamma } }}_k}} \right)
\label{eq:Fusion_q}
\end{equation}
which completes the state update for the $\boldsymbol{\eta}$ partition. This choice forces the conditional term in (\ref{eq:State_Schur}) to its minimum value of zero, thereby ensuring the optimality of the full-state solution in (\ref{eq:MAP_criterion}). The corresponding covariance blocks of the full fusion state are then updated as
\begin{equation}
{{\bf{\hat P}}_{{\bm{\eta \gamma }}}} = {{\bf{\mathord{\buildrel{\lower3pt\hbox{$\scriptscriptstyle\frown$}} \over P} }}_{{\bm{\eta \gamma }}}}{\bf{\mathord{\buildrel{\lower3pt\hbox{$\scriptscriptstyle\frown$}} \over P} }}_{{\bm{\gamma \gamma }}}^{ - 1}{{\bf{\hat P}}_{{\bm{\gamma \gamma }}}}
\label{eq:Fusion_cross_cov}
\end{equation}
\begin{equation}
{{\bf{\hat P}}_{{\bm{\eta \eta }}}} = {{\bf{\mathord{\buildrel{\lower3pt\hbox{$\scriptscriptstyle\frown$}} \over P} }}_{{\bm{\eta \eta }}}} + {{\bf{\mathord{\buildrel{\lower3pt\hbox{$\scriptscriptstyle\frown$}} \over P} }}_{{\bm{\eta \gamma }}}}{\bf{\mathord{\buildrel{\lower3pt\hbox{$\scriptscriptstyle\frown$}} \over P} }}_{{\bm{\gamma \gamma }}}^{ - 1}\left( {{{{\bf{\hat P}}}_{{\bm{\gamma \gamma }}}} - {{{\bf{\mathord{\buildrel{\lower3pt\hbox{$\scriptscriptstyle\frown$}} \over P} }}}_{{\bm{\gamma \gamma }}}}} \right){\bf{\mathord{\buildrel{\lower3pt\hbox{$\scriptscriptstyle\frown$}} \over P} }}_{{\bm{\gamma \gamma }}}^{ - 1}{{\bf{\mathord{\buildrel{\lower3pt\hbox{$\scriptscriptstyle\frown$}} \over P} }}}_{{\bm{\gamma \eta }}}
\label{eq:Fusion_q_cov}
\end{equation}
Equations \eqref{eq:Fusion_q}--\eqref{eq:Fusion_q_cov} complete the full-state recovery. Notably, even though the DPE measurement does not directly observe the $\boldsymbol{\eta}$ state, their posterior estimates are still updated through the prior cross-covariance ${{\bf{\mathord{\buildrel{\lower3pt\hbox{$\scriptscriptstyle\frown$}} \over P} }}_{{\bm{\eta \gamma }}}}$, which propagates the information gained from $\boldsymbol{\gamma}$ to $\boldsymbol{\eta}$. Such an information propagation and feedback process ensures the convergence of attitude and IMU biases within the deep integration system.

In summary, the key distinction between the proposed DPE/INS deep integration and the quasi-deep integration lies in the fact that the system does not produce standalone DPE solutions as intermediate outputs. Rather, it directly performs state and covariance updates via a joint optimization of DPE and INS, which guarantees fusion at the correlation domain. The iterative optimization scheme is well-suited to addressing the nonlinearities inherent in correlation-domain fusion, and the subsequent feedback of the updated state and covariance to the non-PVT states ensures the stability and robustness of the overall system.

\subsection{Summary of the DPE/INS deep integration algorithm}

The estimation procedure derived above is summarized in Algorithm \ref{alg:DPE_INS_deep}. The core step of the overall procedure is the optimization of the DPE/INS joint cost function, which is solved via gradient-based iterative methods, typically the L-M algorithm. No independent DPE solution is produced, thereby achieving a deep fusion of DPE and INS directly in correlation domain. In addition, the selection of the DPE spatial sampling points depends on the INS estimation results and their associated covariances. This aspect will be discussed in detail later. 



\begin{algorithm}[t]
    \caption{DPE/INS deep integration algorithm}
    \renewcommand{\algorithmicrequire}{\textbf{Input:}}
    \renewcommand{\algorithmicensure}{\textbf{Output:}}
    \begin{algorithmic}[1]
        \REQUIRE IMU data ${{\bf U}_K}=\{{{\bf u}_0},\ldots,{{\bf u}_{K-1}}\}$; received IF signal data $s[n]$; IMU and clock noise parameters ${\bf Q}_0$; initial fusion state ${{\bf{\hat x}}_0}$ and covariance ${{\bf{\hat P}}_0}$
        \ENSURE Full fusion estimation results  $\{{{\bf{\hat x}}_k}\}_{k=1}^{K}$ and covariance $\{{{\bf{\hat P}}_k}\}_{k=1}^{K}$ 
        \STATE \(k\leftarrow1\)
        \WHILE{$k < K$}
            \STATE Calculate the predicted full states ${{\bf{\mathord{\buildrel{\lower3pt\hbox{$\scriptscriptstyle\frown$}} \over  x } }}_k}$ through the INS mechanization and clock model using (\ref{eq:INS_derive})\\
            \STATE Obtain the predicted covariance ${{\bf{\mathord{\buildrel{\lower3pt\hbox{$\scriptscriptstyle\frown$}} \over P} }}_k}$ using (\ref{eq:INS_covariance})\\
            \STATE Extract the predicted PVT state ${{\bm{\mathord{\buildrel{\lower3pt\hbox{$\scriptscriptstyle\frown$}} \over \gamma } }}_k}$ from ${{\bf{\mathord{\buildrel{\lower3pt\hbox{$\scriptscriptstyle\frown$}} 
\over x} }}_k}$ and the covariance blocks ${{\bf{\mathord{\buildrel{\lower3pt\hbox{$\scriptscriptstyle\frown$}} \over P} }}_{{\bm{\gamma\gamma}}}}$ , ${{\bf{\mathord{\buildrel{\lower3pt\hbox{$\scriptscriptstyle\frown$}} \over P} }}_{{\bf x} {\bm \gamma}}}$ from ${{\bf{\mathord{\buildrel{\lower3pt\hbox{$\scriptscriptstyle\frown$}} \over P} }}_k}$ as (\ref{eq:P_block})
            \STATE Select the DPE spatial samples $\left\{ {{{{\bm{\tilde \gamma }}}^{1,1}}, \cdots {{{\bm{\tilde \gamma }}}^{1,L}}, \cdots ,{{{\bm{\tilde \gamma }}}^{M,L}}} \right\}$ using the LOS sampling strategy
            \STATE Obtain DPE correlation results ${\Re ^m}\left( {{{{\bm{\tilde \gamma }}}^{m,l}}} \right) $ using (\ref{eq:Correlation_define}) and construct total measurements ${\bf{z}}$ using (\ref{eq:z_define})
           \STATE Prepare the measurement function ${\bf h}\left(\cdot\right)$ and the noise covariance ${{\bf{R}}_k}$
           \STATE Solve the optimization problem (\ref{eq:MAP_PVT}) by L-M method to obtain PVT estimation ${{\bm{\hat \gamma }}_k}$\\
            \STATE Compute the local posterior covariance ${{\bf{\hat P}}_{{\bm{\gamma \gamma }}}}$ for PVT state using (\ref{eq:Fusion_PVT_cov})\\
            \STATE Calculate the remaining states estimation ${{\bm{\hat \eta }}_k}$ using (\ref{eq:Fusion_q}) and combining ${{\bm{\hat \gamma }}_k}$ to recover the full-state posterior estimation ${{\bf{\hat x }}_k}$
            \STATE Calculate the remaining states covariance ${{\bf{\hat P}}_{{\bm{\eta \gamma }}}}$ and ${{\bf{\hat P}}_{{\bm{\eta \eta }}}}$  using (\ref{eq:Fusion_cross_cov})-(\ref{eq:Fusion_q_cov}) and combining ${{\bf{\hat P}}_{{\bm{\gamma \gamma }}}}$ to recover the full-state posterior covariance ${{\bf{\hat P}}_k}$
           \STATE \(k\leftarrow k+1\)
        \ENDWHILE
    \end{algorithmic}
    \label{alg:DPE_INS_deep}
\end{algorithm}

\section{THE PERFORMANCE ANALYSIS AND EVALUATION FOR DPE/INS DEEP INTEGRATION}

\subsection{The Cramér-Rao Lower Bound analysis for DPE/INS deep integration}

To establish a theoretical benchmark for the proposed DPE/INS deep integration framework, we analyze the Posterior Cramér-Rao Lower Bound (PCRLB) for the fusion system. The PCRLB is adopted due to its recursive formulation, which aligns seamlessly with the sequential prediction-update structure of the proposed DPE/INS deep integration framework. For the state ${{{\bf{x}}_k}}$ at epoch $k$, the estimation error covariance satisfies:
\begin{equation}
{{\bf{\hat P}}_k} = E\left[ {\left( {{{\bf{x}}_k} - {{{\bf{\hat x}}}_k}} \right){{\left( {{{\bf{x}}_k} - {{{\bf{\hat x}}}_k}} \right)}^T}} \right] \succeq {\bf{I}}_k^{ - 1}
\label{eq:CRB_define}
\end{equation}
where ${{\bf{I}}_k}$ is the Fisher Information Matrix (FIM). The recursive computation of the FIM for a discrete-time nonlinear system is given as \cite{ref:PCRLB} 
\begin{equation}
{{\bf{I}}_k} = {\bf{H}}_{k - 1}^{22} - {\bf{H}}_{k - 1}^{21}{\left( {{{\bf{I}}_{k - 1}} + {\bf{H}}_{k - 1}^{11}} \right)^{ - 1}}{\bf{H}}_{k - 1}^{12}
\label{eq:PCRB_define}
\end{equation}
where the matrices ${\bf{H}}_{k - 1}^{11}$, ${\bf{H}}_{k - 1}^{12}$, ${\bf{H}}_{k - 1}^{21}$, and ${\bf{H}}_{k - 1}^{22}$ are defined based on the second-order derivatives of the state transition and measurement log-likelihoods.

To derive the recursive FIM, we first compute the second-order derivatives of the log-likelihood functions. For the state transition model $p\left( {{{\bf{x}}_k}|{{\bf{x}}_{k - 1}}} \right) \sim {\mathcal{N}}\left( {{{\bf{F}}_{k - 1}}{{\bf{x}}_{k - 1}},{{\bf{Q}}_{k - 1}}} \right)$, the matrices ${\bf{H}}_{k - 1}^{11}$, ${\bf{H}}_{k - 1}^{12}$, and ${\bf{H}}_{k - 1}^{21}$ are obtained as: 
\begin{equation}
{\bf{H}}_{k - 1}^{11} =  - {\rm E}\left[ {\Delta _{{{\bf{x}}_{k - 1}}}^{{{\bf{x}}_{k - 1}}}\ln p\left( {{{\bf{x}}_k}|{{\bf{x}}_{k - 1}}} \right)} \right] = {\bf{F}}_{k - 1}^T{\bf{ Q}}_{k - 1}^{ - 1}{{\bf{F}}_{k - 1}}
\label{eq:H11}
\end{equation}
\begin{equation}
{\bf{H}}_{k - 1}^{12} =  - {\rm E}\left[ {\Delta _{{{\bf{x}}_{k - 1}}}^{{{\bf{x}}_k}}\ln p\left( {{{\bf{x}}_k}|{{\bf{x}}_{k - 1}}} \right)} \right] =  - {\bf{F}}_{k - 1}^T{\bf{ Q}}_{k - 1}^{ - 1}
\label{eq:H12}
\end{equation}
\begin{equation}
{\bf{H}}_{k - 1}^{21} =  - {\rm E}\left[ {\Delta _{{{\bf{x}}_k}}^{{{\bf{x}}_{k - 1}}}\ln p\left( {{{\bf{x}}_k}|{{\bf{x}}_{k - 1}}} \right)} \right] =  - {\bf{ Q}}_{k - 1}^{ - 1}{{\bf{F}}_{k - 1}}
\label{eq:H21}
\end{equation}
where the symbol $\Delta _{\bf{a}}^{\bf{b}} \buildrel \Delta \over = {\partial ^2}/\partial {\bf{a}}\partial {{\bf{b}}^T}$ denotes the second-order partial derivative operator. For the measurement model $p\left( {{{\bf{z}}_k}|{{\bf{x}}_k}} \right) \sim {\mathcal{N}}\left( {{\bf{h}}\left( {{\bf{S}}_{\bm{\gamma }}}{{{\bf{x}}_k}} \right),{{\bf{R}}_k}} \right)$, and combining with the state transition contribution, we have
\begin{equation}
\begin{aligned}
{\bf{H}}_{k - 1}^{22} =& - {\rm E}\left[ {\Delta _{{{\bf{x}}_k}}^{{{\bf{x}}_k}}\ln p\left( {{{\bf{x}}_k}|{{\bf{x}}_{k - 1}}} \right)} \right] - {\rm E}\left[ {\Delta _{{{\bf{x}}_k}}^{{{\bf{x}}_k}}\ln p\left({{{\bf{z}}_k}|{{\bf{x}}_k}} \right)} \right]\\
 =& {\bf{ Q}}_{k - 1}^{ - 1} + {\bf{S}}_{\bm{\gamma }}^T{\bf{J}}_k^T{\bf{R}}_k^{ - 1}{{\bf{J}}_k}{{\bf{S}}_{\bm{\gamma }}}
\end{aligned}
\label{eq:H22}
\end{equation}
Substituting these four $\bf H$ matrices into the general PCRLB recursion (\ref{eq:PCRB_define}) and applying the Woodbury matrix identity \cite{ref:Woodbury}, we derive the final recursive form of the FIM for the proposed DPE/INS deep integration framework
\begin{equation}
\begin{aligned}
{{\bf{I}}_k} =& {\bf{ Q}}_{k - 1}^{ - 1} + {\bf{S}}_{\bm{\gamma }}^T{\bf{J}}_k^T{\bf{R}}_k^{ - 1}{{\bf{J}}_k}{{\bf{S}}_{\bm{\gamma }}}\\
&-{\bf{ Q}}_{k - 1}^{ - 1}{{\bf{F}}_{k - 1}}{\left( {{{\bf{I}}_{k - 1}} + {\bf{F}}_{k - 1}^T{\bf{ Q}}_{k - 1}^{ - 1}{{\bf{F}}_{k - 1}}} \right)^{ - 1}}{\bf{F}}_{k - 1}^T{\bf{Q}}_{k - 1}^{ - 1}\\
=& \underbrace {{{\left( {{{\bf{F}}_{k - 1}}{\bf{I}}_{k - 1}^{ - 1}{\bf{F}}_{k - 1}^T + {{{\bf{ Q}}}_{k - 1}}} \right)}^{ - 1}}}_{INS:{{{\bf{\mathord{\buildrel{\lower3pt\hbox{$\scriptscriptstyle\frown$}} 
\over I} }}}_k}} + \underbrace {{{\bf{S}}_{\bm{\gamma }}^T{\bf{J}}_k^T{\bf{R}}_k^{ - 1}{{\bf{J}}_k}{{\bf{S}}_{\bm{\gamma }}}}}_{DPE:{{{\bf{\mathord{\buildrel{\lower3pt\hbox{$\scriptscriptstyle\smile$}} 
\over I} }}}_k}}
\end{aligned}
\label{eq:FIM_fusion}
\end{equation}
This decomposition reveals that the total FIM at epoch $k$ is the sum of two parts

\begin{itemize}
  \item INS Contribution: \\${{\bf{\mathord{\buildrel{\lower3pt\hbox{$\scriptscriptstyle\frown$}} 
\over I} }}_k} = {\left( {{{\bf{F}}_{k - 1}}{\bf{I}}_{k - 1}^{ - 1}{\bf{F}}_{k - 1}^T + {{{\bf{ Q}}}_{k - 1}}} \right)^{ - 1}}$, this term represents the information propagated from the previous epoch through the INS prediction. It encapsulates the time coherence of the IMU errors via the equivalent noise covariance ${{{{\bf{ Q}}}_{k - 1}}}$.
  \item DPE Contribution: ${{\bf{\mathord{\buildrel{\lower3pt\hbox{$\scriptscriptstyle\smile$}} 
\over I} }}_k} = {{\bf{S}}_{\bm{\gamma }}^T{\bf{J}}_k^T{\bf{R}}_k^{ - 1}{{\bf{J}}_k}{{\bf{S}}_{\bm{\gamma }}}}$, this term captures the information provided by the DPE measurements. It encodes the spatial coherence structure of the correlation function via ${{{\bf{R}}_k}}$. The selection matrix 
${{{\bf{S}}_{\bm{\gamma }}}}$ explicitly indicates that the DPE measurement information directly affects only the PVT state $\bm \gamma$, while providing no direct information contribution to the remaining $\bm \eta$ state.
\end{itemize}

The additive nature of the information matrix in (\ref{eq:FIM_fusion}) demonstrates that the proposed fusion framework achieves optimal information integration, analogous to the update step shown in Algorithm \ref{alg:DPE_INS_deep}. This provides a theoretical foundation for the subsequent steady-state performance analysis.

\subsection{The steady-state performance analysis}

The recursive PCRLB in (\ref{eq:FIM_fusion}) describes the instantaneous information propagation of the proposed DPE/INS deep integration framework. To further reveal the long-term performance, a local steady-state condition is considered. In this condition, the system parameters are assumed to remain approximately constant within a short time window. Therefore, the fusion covariance and INS prediction covariance converge to ${{\bf{\hat P}}_\infty}$ and ${{\bf{\mathord{\buildrel{\lower3pt\hbox{$\scriptscriptstyle\frown$}}\over P}}}_\infty}$ respectively.

Based on the recursive FIM expression in (\ref{eq:FIM_fusion}), the steady-state covariance satisfies the following relationship
\begin{equation}
{{\bf{\mathord{\buildrel{\lower3pt\hbox{$\scriptscriptstyle\frown$}} \over P} }}_\infty } = {\bf{F}}{{\bf{\hat P}}_\infty }{{\bf{F}}^T} + {\bf{Q}}
\label{eq:steady_P_ins}
\end{equation}
\begin{equation}
{{\bf{\hat P}}_\infty } = {\left( {{\bf{\mathord{\buildrel{\lower3pt\hbox{$\scriptscriptstyle\frown$}} 
\over P} }}_\infty ^{ - 1} + {\bf{\mathord{\buildrel{\lower3pt\hbox{$\scriptscriptstyle\smile$}} 
\over I} }}} \right)^{ - 1}}
\label{eq:steady_P_fusion}
\end{equation}
where $\bf F$ describes the dynamic characteristic of the vehicle, and \(\mathbf{Q}\) encapsulates the IMU and clock noise parameters. The FIM \({\bf{\mathord{\buildrel{\lower3pt\hbox{$\scriptscriptstyle\smile$}} 
\over I} }}\) represents the information contribution from the DPE measurements, which is determined by the satellite geometry, carrier-to-noise ratios, and the spatial sampling strategy.

Since the DPE measurements directly observe only the PVT state $\bm \gamma$, our analysis focuses on this subset. Moreover, based on the distinct measurement characteristics of DPE, the PVT state $\bm \gamma$ can be further partitioned into two separate blocks
\begin{equation}
{{\bm{\gamma }}_{\bf{p}}} = {\left[ {\begin{array}{*{20}{c}}
{{{\bf{p}}^T}}&{c\delta t}
\end{array}} \right]^T} = {{\bf{S}}_{{{\bm{\gamma }}_{\bf{p}}}}}{\bf{x}}
\label{eq:P_block_define}
\end{equation}
\begin{equation}
{{\bm{\gamma }}_{\bf{v}}} = {\left[ {\begin{array}{*{20}{c}}
{{{\bf{v}}^T}}&{c\delta \dot t}
\end{array}} \right]^T} = {{\bf{S}}_{{{\bm{\gamma }}_{\bf{v}}}}}{\bf{x}}
\label{eq:V_block_define}
\end{equation}
which correspond to the position-clock bias and velocity-clock drift components, respectively. Let \(\boldsymbol{\chi} \in \{\mathbf{p}, \mathbf{v}\}\) denote a generic block index. The corresponding covariance and information sub-matrices are denoted by subscript \(\boldsymbol{\chi}\), e.g., \({{\bf{\hat P}}_{\boldsymbol{\chi}}}\) and \({{\bf{\mathord{\buildrel{\lower3pt\hbox{$\scriptscriptstyle\smile$}} \over I} }}_{\boldsymbol{\chi}}}\). Specifically, the available DPE information can be written as \cite{ref:LOS sampling}
\begin{equation}
{{\bf{\mathord{\buildrel{\lower3pt\hbox{$\scriptscriptstyle\smile$}} \over I} }}_{{\bm{\chi }}}} = {F_{\bm{\chi }}}({\alpha _{\bm{\chi }}},{\beta _{\bm{\chi }}}){{\bf{B}}_{\bm{\chi }}}
\label{eq:information_provided}
\end{equation}
\begin{equation}
{{\bf{B}}_{\bm{\chi }}} = {s_{\bm{\chi }}}{{\bf{G}}^T}{\bf{\Omega G}}
\label{eq:information_provided_matrix}
\end{equation}
Here ${\bf G}$ is the satellite geometry matrix, and ${\bf \Omega}={\rm diag}(C/N_0^1,\cdots,C/N_0^M)$ collects the carrier-to-noise ratios of all visible satellites. The scale factors are $s_{\bf p}=f_c^2T_c/c^2$ for the code-delay related block and $s_{\bf v}=f_L^2T_c^3/c^2$ for the Doppler related block, where $f_c$ and $f_L$ represent the code frequency and carrier frequency, respectively. Therefore, the influence of DPE spatial sampling on the steady-state PCRLB is compressed into the scalar factor $F_{\bm \chi}$, while the satellite geometry and signal strength are fixed as reference values in ${{\bf{B}}_{\bm{\chi }}}$. 

The normalized scale factor $F_{{\bm \chi}}$ characterizes the influence of sampling parameters on FIM, including the number of sampling intervals $\alpha_{{\bm \chi}}$ and sampling step $\beta_{{\bm \chi}}$ operated along the LOS direction
\begin{equation}
{F_{\bm{\chi }}}({\alpha _{\bm{\chi }}},{\beta _{\bm{\chi }}})=
\left( {\bf J}_{{\bm \chi}} \right)^{T}
\left( {\bf R}_{{\bm \chi}} \right)^{-1}
{\bf J}_{{\bm \chi}}<{F_{\bm{\chi }}^{\rm max}}
\label{eq:F0}
\end{equation}
The normalized covariance matrix ${\bf R}_{{\bm \chi}}$ and normalized Jacobian vector ${\bf J}_{{\bm \chi}}$ are \cite{ref:LOS sampling}

\begin{equation}
\left[{\bf R}_{{\bm \chi}}\right]_{i,j}
=
\begin{cases}
\max\left(0,1-\frac{2\beta_{\bf p}f_c}{c}\left|j-i\right|\right),
& {\bm \chi}={\bf p}\\[1.0ex]
\operatorname{sinc}\left(\frac{2\beta_{\bf v}f_LT_c}{c}\left|j-i\right|\right),
& {\bm \chi}={\bf v}
\end{cases}
\label{eq:R_LOS_chi}
\end{equation}

\begin{equation}
\left[{\bf J}_{{\bm \chi}}\right]_{l}
=
\begin{cases}
\operatorname{sign}\left(l-\frac{\alpha_{\bf p}}{2}-1\right),
& {\bm \chi}={\bf p}\\[1.0ex]
\operatorname{sinc}^{\prime}\left[
\frac{2\beta_{\bf v}f_LT_c}{c}
\left(l-\frac{\alpha_{\bf v}}{2}-1\right)
\right],
& {\bm \chi}={\bf v}
\end{cases}
\label{eq:J_LOS_chi}
\end{equation}
where $[\cdot]_{i,j}$ denotes the $(i,j)$-th matrix element, $[\cdot]_{l}$ denotes the $l$-th vector element. It can be observed that the normalized factor $F_{{\bm \chi}}$ is determined by sampling parameters $\alpha_{{\bm \chi}}$ and $\beta_{{\bm \chi}}$. The factor ${F_{\bm{\chi }}^{\rm max}}$ represents the upper bound of the information content determined by the GNSS signal itself and is independent of the LOS sampling parameters.
\begin{equation}
F_{{\bf p}}^{\rm max}
= 8\pi^{2}\left(B_n/f_c\right)^{2}; \quad
F_{{\bf v}}^{\rm max}
= 2\pi^{2}/3
\label{eq:F_max}
\end{equation}
where $B_n$ is the root-mean-square bandwidth. Reasonable LOS sampling parameters can approach this upper bound, which also means more sufficient utilization of information.

It should be noted that the system dynamics parameters in 
$\bf F$, IMU and clock parameters in $\bf Q$, and the satellite geometry and signal strength parameters in ${{\bf{B}}_{\bm{\chi }}}$ are all inherent to the system and cannot be modified. The only adjustable items are the sampling parameters $\alpha_{{\bm \chi}}$ and $\beta_{{\bm \chi}}$ in the normalized scale factor $F_{{\bm \chi}}$. Consequently, for a given set of system parameters and a specified performance target, it is necessary to identify the conditions under which the sampling parameters fulfill the requirement.

Let ${{\bf{\hat P}}^\star}$ be a desired steady-state covariance for DPE/INS fusion. For each block ${\bm \chi} \in\{{\bf p},{\bf v}\}$, the  amount of DPE information required to achieve the target covariance ${{\bf{\hat P}}^\star}$ is derived from (\ref{eq:steady_P_ins})-(\ref{eq:steady_P_fusion})
\begin{equation}
{\bf{\mathord{\buildrel{\lower3pt\hbox{$\scriptscriptstyle\smile$}} 
\over I} }}_{\bm{\chi }}^ \star  = {\left( {{\bf{\hat P}}_{\bm{\chi }}^ \star } \right)^{ - 1}} - {\left[ {{{\bf{S}}_{{{\bm{\gamma }}_{\bm{\chi }}}}}\left( {{\bf{F\hat P}}^ \star {{\bf{F}}^T} + {\bf{Q}}} \right){\bf{S}}_{{{\bm{\gamma }}_{\bm{\chi }}}}^T} \right]^{ - 1}}
\label{eq:DPE_information_required}
\end{equation}
For the available DPE information to satisfy the target steady-state
performance in block $\bm \chi$, the following matrix inequality should hold:
\begin{equation}
{{\bf{\mathord{\buildrel{\lower3pt\hbox{$\scriptscriptstyle\smile$}} \over I} }}_{{\bm{\chi }}}} = {F_{\bm{\chi }}}({\alpha _{\bm{\chi }}},{\beta _{\bm{\chi }}}){{\bf{B}}_{\bm{\chi }}} \succeq {\bf{\mathord{\buildrel{\lower3pt\hbox{$\scriptscriptstyle\smile$}} 
\over I} }}_{\bf{\chi }}^ \star
\label{eq:I_constraint}
\end{equation}
Equivalently, the minimum required LOS information factor is given by the largest generalized eigenvalue
\begin{equation}
F_{\bm{\chi }}^ \star = \mathop {\max }\limits_{\left\| {{{\bf{a}}_{\bm{\chi }}}} \right\| = 1} \frac{{{\bf{a}}_{\bm{\chi }}^T{\bf{\mathord{\buildrel{\lower3pt\hbox{$\scriptscriptstyle\smile$}} 
\over I} }}_{\bm{\chi }}^ * {{\bf{a}}_{\bm{\chi }}}}}{{{\bf{a}}_{\bm{\chi }}^T{{\bf{B}}_{\bm{\chi }}}{{\bf{a}}_{\bm{\chi }}}}}
\label{eq:F_required}
\end{equation}
The vector ${\bf a}_{\bm{\chi}}$ is a projection direction in the four-dimensional
block ${{\bm \gamma}_{\bm \chi}}$, rather than an additional system variable. When ${\bf B}_{\bm \chi}$ is rank deficient, the generalized eigenvalue condition is evaluated in the observable subspace of ${\bf B}_{\bm \chi}$; otherwise, the desired covariance cannot be guaranteed by the available system parameters.

Finally, the required sampling parameters condition for DPE is obtained as
\begin{equation}
\begin{aligned}
{\cal S}_{\bm \chi}^{{\rm{info}}} = &\{ ({\alpha _{\bm\chi}},{\beta _{\bm\chi}})\;|\;{\alpha _{\bm\chi}} \in 2{{\rm N}^ + },{\beta _{\bm\chi}} > 0,\\
&{F_{\bm\chi} }({\alpha _{\bm\chi}},{\beta _{\bm\chi}}) \ge F_{\bm\chi} ^ \star ({\bf{F}},{\bf{Q}},{{\bf{B}}_{\bm\chi} },{\bf{\hat P}}^ \star )\}
\end{aligned}
\label{eq:information condition}
\end{equation}
where $\bf F$, $\bf Q$, and ${\bf B}_{\bm \chi}$ are intrinsic system parameters, denoting the system dynamics, the IMU and clock stochastic error, and the satellite geometry together with signal strengths, respectively, while ${\bf{\hat P}}^ \star $ represents the desired steady-state covariance of the fusion system.

\subsection{The INS aiding on DPE spatial sampling}

The previous subsection establishes the information-sufficiency condition on the DPE spatial sampling parameters. Specifically, for each DPE-observable block $\bm \chi \in \{\mathbf{p}, \mathbf{v}\}$, the normalized factor $F_{\bm \chi}(\alpha_{\bm\chi}, \beta_{\bm\chi})$ must be sufficiently large to meet the target steady-state performance. This condition corresponds to the ideal case where the sampling points completely cover the main peak of the correlation function. In practical scenarios, the sampling center is perturbed by errors. To ensure that the sampling points adequately cover the main peak, the INS prediction information is exploited to guide the design of the DPE sampling strategy. 

Given the INS prediction information, the INS-predicted state can be adopted as the center of the LOS sampled points for DPE receiver. Accordingly, the set of sampling points for the $m$-th satellite is constructed as \cite{ref:LOS sampling}
\begin{equation}
\left\{ {{\bm{\tilde \gamma }}_{\bm{\chi }}^{m,l}} \right\}_{l = 1}^{{\alpha _{\bm{\chi }}} + 1} = \{ {{\bm{\mathord{\buildrel{\lower3pt\hbox{$\scriptscriptstyle\frown$}} 
\over \gamma } }}_{\bm{\chi }}} + \lambda _{\bm{\chi }}^l{\beta _{\bm{\chi }}}{{\bf{\bar u}}^m}|\lambda _{\bm{\chi }}^l = l - \frac{{{\alpha _{\bm{\chi }}}}}{2} - 1\} 
\label{eq:ins_aided_sampling_points}
\end{equation}
\begin{equation}
{{\bf{\bar u}}^m} = {\left[ {\begin{array}{*{20}{c}}
{{{\left( {{\bf{p}}_s^m - {\bf{p}}} \right)}^T}/\left\| {{\bf{p}}_s^m - {\bf{p}}} \right\|}&{ - 1}
\end{array}} \right]^T}
\label{eq:ins_aided_extended_los}
\end{equation}
where ${{\bf{\bar u}}^m}$ denotes the extended LOS vector for the $m$-th satellite, which defines the spatial sampling orientation; ${{\bm{\mathord{\buildrel{\lower3pt\hbox{$\scriptscriptstyle\frown$}} 
\over \gamma } }}_{\bm{\chi }}}$ is the INS-predicted PVT state, which serves as the sampling center; and $l=1,\cdots,\alpha_{\bm \chi}+1$ sequentially indexes the sampling points, while $\beta_{\bm \chi}$ governs the spacing between adjacent points. 

The actual correlation peak is shifted from the sampling center by the combined uncertainty from both the INS prediction and the current DPE estimation. Hence, the total uncertainty that determines the possible shift of the correlation peak is the sum of two contributions. The INS prediction uncertainty is characterized by the INS prediction covariance ${\bf{\mathord{\buildrel{\lower3pt\hbox{$\scriptscriptstyle\frown$}} \over P} }}_{{\bm{\chi}}}$. The current DPE estimation covariance can be approximately predicted by the DPE information matrix shown as
\begin{equation}
{{\bf{\mathord{\buildrel{\lower3pt\hbox{$\scriptscriptstyle\smile$}} 
\over P} }}_{\bm{\chi }}} = {\left( {F_{\bm{\chi }}^ * {s_{\bm{\chi }}}{{\bf{G}}^T}{\bf{\Omega G}}} \right)^{ - 1}}
\label{eq:DPE_covariance}
\end{equation}
The projection of these combined uncertainties onto the $m$-th LOS direction is
\begin{equation}
{\left( {\sigma _{\bm{\chi }}^m} \right)^2} = E\left[ {{{\left( {\xi _{\bm{\chi }}^m} \right)}^2}} \right] = \frac{1}{4}{\left( {{\bf{\bar u}}^m}  \right)^T}\left( {{{{\bf{\mathord{\buildrel{\lower3pt\hbox{$\scriptscriptstyle\frown$}} 
\over P} }}}_{\bm{\chi }}} + {{{\bf{\mathord{\buildrel{\lower3pt\hbox{$\scriptscriptstyle\smile$}} 
\over P} }}}_{\bm{\chi }}}} \right){{\bf{\bar u}}^m}
\label{eq:ins_aided_projected_uncertainty}
\end{equation}
Here $\xi _{\bm \chi}^m$ measures the displacement of the actual correlation peak relative to the sampling center from INS prediction, and $\sigma_{\bm\chi}^{m}$ is its standard deviation. The coefficient derives The factor 1/4 of (\ref{eq:ins_aided_projected_uncertainty}) follows from ${\left\| {{\bf{\bar u}}^m} \right\|^4}=4$. 

The ideal main-peak half width is $\rho_{\bf p}=c/(2f_c)$ for the position-clock bias block and $\rho_{\bf v}=c/(2f_LT_c)$ for the velocity-clock drift block. To preserve the effective gradient information of the correlation function, the sampled points should cover the main peak around the true correlation peak. For a required coverage probability $p_c$, the following condition is imposed:
\begin{equation}
{\rm P}
\left(
\left|\xi_{\bm\chi}^{m}\right|
\le
\frac{\alpha_{\bm\chi}\beta_{\bm\chi}}{2}
-\rho_{\bm\chi}
\right)
\ge
p_c
\label{eq:ins_aided_probabilistic_coverage}
\end{equation}
Using the Gaussian approximation in (\ref{eq:ins_aided_projected_uncertainty}),
the probability condition (\ref{eq:ins_aided_probabilistic_coverage}) leads to the deterministic INS-aided uncertainty-coverage constraint
\begin{equation}
\begin{aligned}
{\cal S}_{\bm\chi}^{{\rm{cov}}} = \{ ({\alpha _{\bm\chi }},{\beta _{\bm\chi}})\;|\;{\alpha _{\bm\chi}} \in 2{{\rm N}^ + },{\beta _{\bm\chi }} > 0,\\
\frac{{{\alpha _{{\bm{\chi }}}}{\beta _{{\bm{\chi }}}}}}{2} \ge {\rho _{\bm{\chi }}} + {z_{{p_c}}}\mathop {\max }\limits_{m = 1, \cdots ,M} \sigma _{{\bm{\chi }}}^m\}
\end{aligned}
\label{eq:ins_aided_coverage_constraint}
\end{equation}
where $z_{p_c}=\Phi^{-1}((1+p_c)/2)$ and $\Phi(\cdot)$ denotes the standard
normal cumulative distribution function. Equation (\ref{eq:ins_aided_coverage_constraint}) shows that a more accurate INS prediction reduces the projected uncertainty $\sigma_{\bm\chi}^{m}$ and therefore allows a shorter sampling segment. Conversely, when the INS prediction uncertainty increases, the sampling region should be expanded to avoid missing the correlation peak.

Combining the steady-state information requirement and the INS-aided coverage requirement, the feasible sampling set for block $\bm\chi$ is
\begin{equation}
\left(\alpha_{\bm\chi},\beta_{\bm\chi}\right)
\in
{\mathcal S}_{\bm\chi}^{\rm info}
\cap
{\mathcal S}_{\bm\chi}^{\rm cov}
\label{eq:ins_aided_joint_feasible_set}
\end{equation}
It is worth noting that the constraint ${\mathcal S}_{\bm\chi}^{\rm info}$ is time-invariant, as it depends solely on the steady-state system parameters and the specified performance target. By contrast, the constraint ${\mathcal S}_{\bm\chi}^{\rm cov}$ is time-varying, since it is evaluated online using the current INS prediction covariance and the estimated DPE covariance. Among all feasible selections, a practical choice is to minimize the number of correlation evaluations:
\begin{equation}
\left(\alpha_{\bm\chi}^{\star},\beta_{\bm\chi}^{\star}\right)
=
\mathop{\arg\min}\limits_{
\left(\alpha_{\bm\chi},\beta_{\bm\chi}\right)
\in
{\mathcal S}_{\bm\chi}^{\rm info}
\cap
{\mathcal S}_{\bm\chi}^{\rm cov}}
\left(\alpha_{\bm\chi}+1\right)
\label{eq:ins_aided_sampling_design}
\end{equation}
Thus, the steady-state analysis provides the information sufficiency constraint, whereas the INS prediction provides the uncertainty-coverage constraint. The former determines whether the sampling parameters retain enough DPE information to satisfy the target steady-state performance, while the latter determines whether the actual sampling segment is centered and expanded appropriately according to the current INS prediction uncertainty.

\section{SIMULATION AND EXPERIMENT VERIFICATION}

\subsection{The PCRLB and steady-state performance analysis}

To verify the recursive PCRLB and the additive FIM structure in (\ref{eq:FIM_fusion}), a 60-s vehicle trajectory consisting of two 18-s turning segments and two 12-s straight segments is designed for simulation. The GNSS coherent integration time is set to \(T_c=20\) ms. Eight GPS L1 satellites are selected, with an initial $\text{GDOP} = \sqrt{\text{tr}\left(({\bf G}^T{\bf G})^{-1}\right)} = 2.08$. The \(C/N_0\) of all satellite channels is reduced from 45 to 40 and then to 35 dB-Hz, with each level maintained for 20 s. For INS parameters, the bias stability and noise density of gyroscopes are set as ${{\bm{\sigma }}_{rg}}={{0.01^\circ }/{\rm{s}}}$ and ${{\bm{\sigma }}_{g}}=0.2^\circ/s/ {\sqrt {Hz} }$, respectively, while ${{\bm{\sigma }}_{ra}}=16{\rm{mg}}$ and ${{\bm{\sigma }}_{a}}={{1.3{\rm{mg}}} / {\sqrt {\rm{Hz}} }}$ for accelerometers bias stability and noise density. The PCRLB and FIM curves are obtained by (\ref{eq:FIM_fusion}) using the above parameter settings, while the Monte Carlo root-mean-square error (RMSE) curves are independently obtained from 200 full-trajectory trials under the same cases.

\begin{figure}[ht]
    \centering
    \includegraphics[width=1\linewidth]{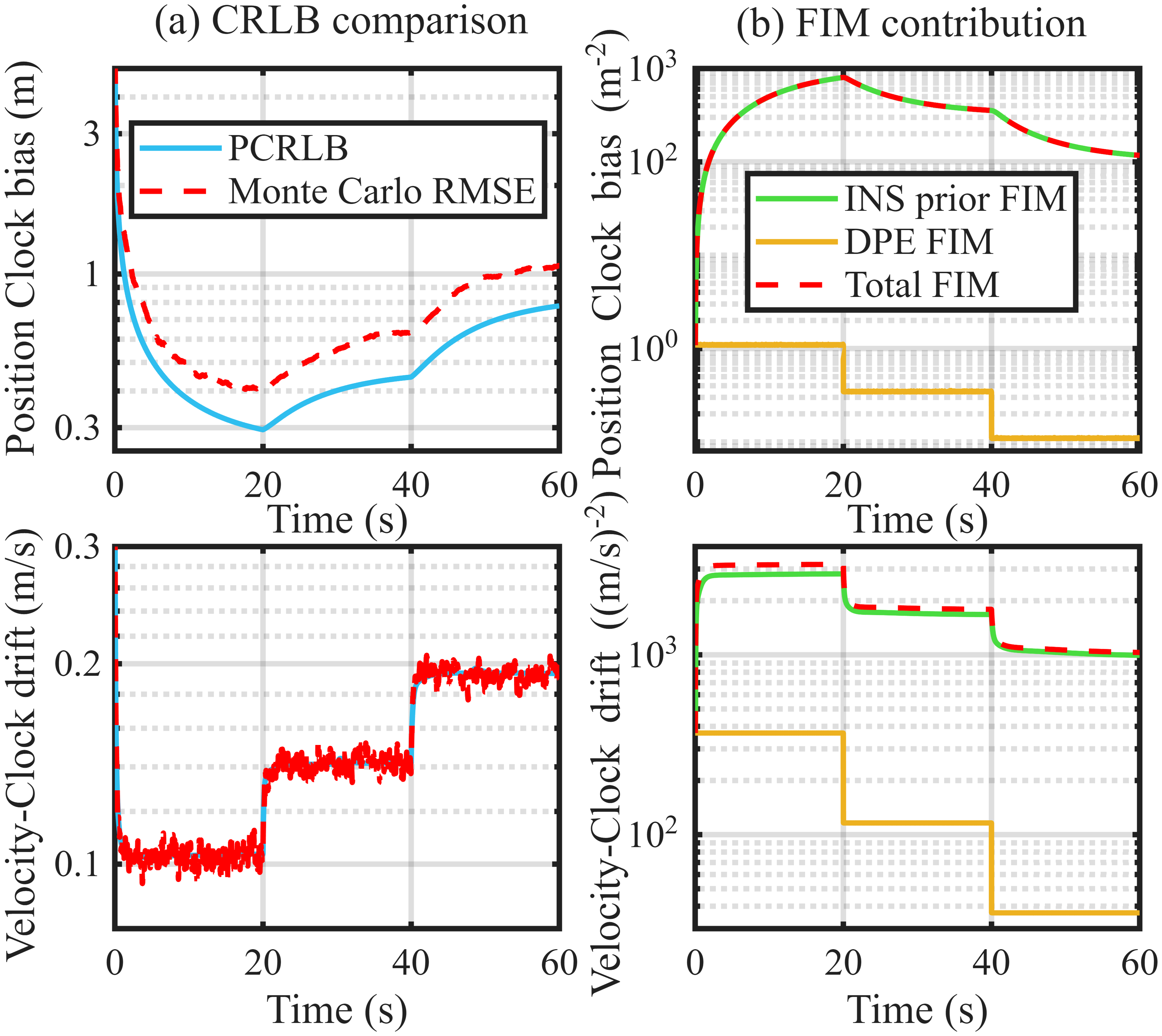}
    \caption{The PCRLB and Monte Carlo RMSE results for DPE/INS deep integration, combined with FIM contribution analysis}
    \label{fig:PCRLB}
\end{figure}

As shown in Fig. \ref{fig:PCRLB}a, the position-clock bias RMSE remains consistently above the corresponding PCRLB, whereas the velocity-clock drift RMSE fluctuates closely around its theoretical bound. For position-clock bias block, the stable PCRLB and Monte Carlo RMSE are respectively 0.30 m and 0.40 m at 45 dB-Hz stage, 0.45 m and 0.63 m at 40 dB-Hz stage, and 0.78 m and 1.07 m at 35 dB-Hz stage, while for velocity-clock drift block, the stable PCRLB and Monte Carlo RMSE are nearly consistent, showing 0.103 m/s, 0.137 m/s, and 0.191 m/s for three stages. This closer agreement for velocity-clock bias block indicates that the Doppler correlation measurements provide a more sensitive and locally better-conditioned constraint on velocity and clock drift.

The FIM decomposition in Fig. \ref{fig:PCRLB}b provides the information-domain explanation for the above phenomenon from a different perspective. For the position-clock bias block, the INS prior FIM almost coincides with the total FIM, indicating that the instantaneous DPE contribution is small compared with the information accumulated through INS propagation. For the velocity-clock drift block, the separation between the INS prior and total FIM curves is clearly visible, showing that the DPE contribution occupies a substantially larger proportion of the total information. On the other hand, the long-term total FIM exhibits three distinct stages corresponding to three \(C/N_0\) levels for both estimation blocks. This indicates that the long-term total information of the deeply integrated system is ultimately governed by the property and quality of the GNSS signal.

\begin{figure}[ht]
    \centering
    \includegraphics[width=1\linewidth]{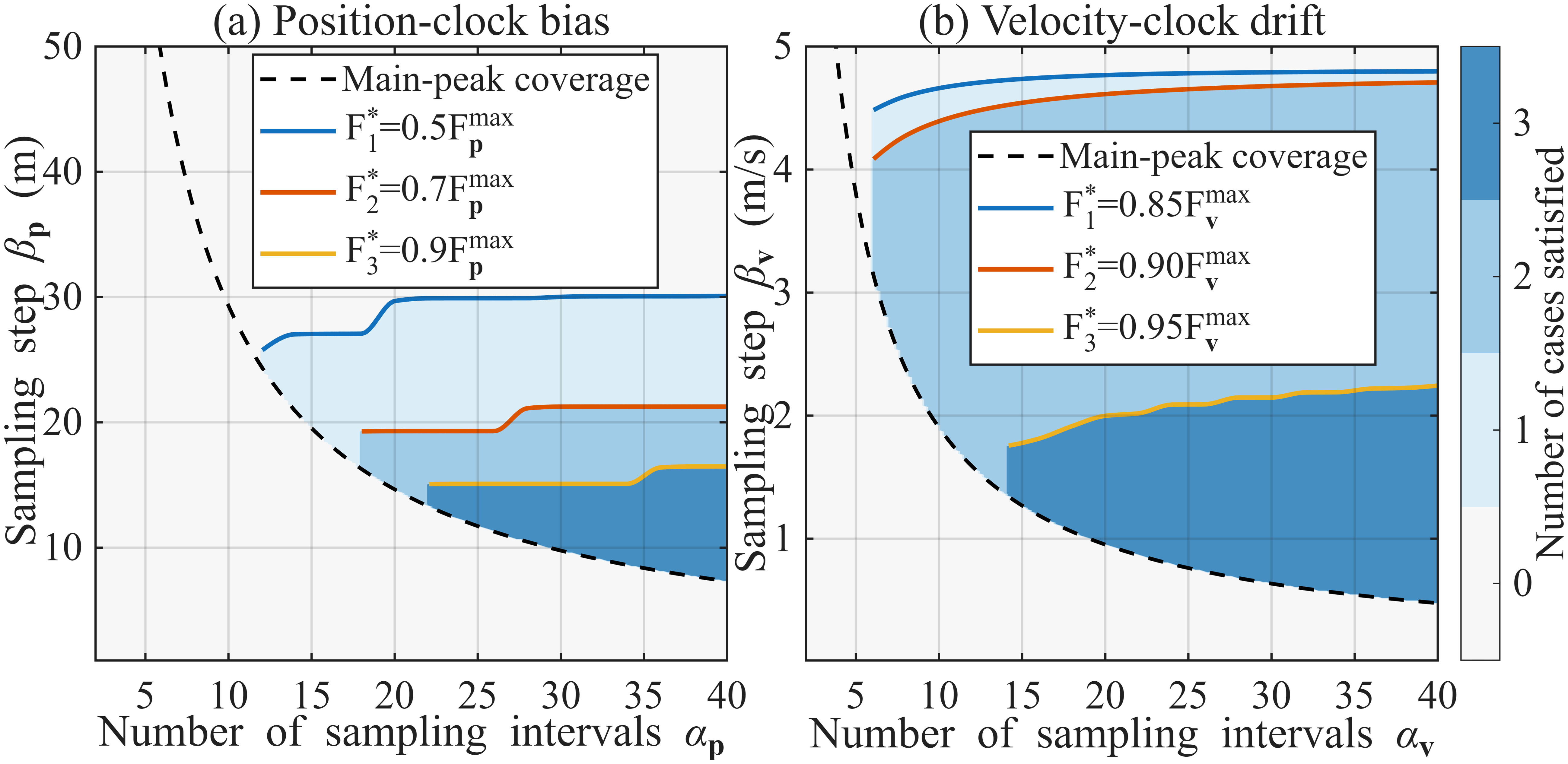}
    \caption{The feasible region of sampling parameter design under different steady-state performance requirements}
    \label{fig:Steady}
\end{figure}

The steady-state performance analysis is performed using a static receiver and the same eight-satellite geometry, with \(C/N_0\) fixed as 40 dB-Hz. The other system parameters remain the same as the Monte Carlo simulation. Fig. \ref{fig:Steady} illustrates the feasible region of sampling parameter design under different steady-state performance requirements. The main-peak coverage curve corresponds to the constraint \(\alpha_{\bm \chi}\beta_{\bm \chi}\geq2\rho_{\bm \chi}\). Three attainable information factors are selected as \(0.5F_{\bf p}^{\max}\), \(0.7F_{\bf p}^{\max}\), and \(0.9F_{\bf p}^{\max}\) for the position block and \(0.85F_{\bf v}^{\max}\), \(0.90F_{\bf v}^{\max}\), and \(0.95F_{\bf v}^{\max}\) for the velocity block. Their corresponding equivalent total standard deviations \(\sqrt{\operatorname{tr}({\bf P}_{\bm \chi}^\star)}\), are 2.41 m, 2.21 m, and 2.08 m, and 0.183 m/s, 0.180 m/s, and 0.177 m/s, respectively, where ${\bf P}_{\bm \chi}^\star$ stands for the steady-state covariance requirements.

For the position-clock-bias block, the minimum numbers of sampling intervals for \(0.5F_{\bf p}^{\max}\), \(0.7F_{\bf p}^{\max}\), and \(0.9F_{\bf p}^{\max}\) targets are  \(\alpha_{\bf p}=12,18,22\). For the velocity-clock-drift block, the \(0.85F_{\bf v}^{\max}\) and \(0.90F_{\bf v}^{\max}\) targets are feasible from \(\alpha_{\bf v}=6\), whereas the \(0.95F_{\bf v}^{\max}\) target requires at least \(\alpha_{\bf v}=14\). Consequently, the feasible region shifts toward a larger number of sampling intervals and smaller step sizes as the information demand increases, or equivalently, as the steady-state performance requirements become stricter. Moreover, compared with the position–clock bias block, the velocity–clock drift block requires fewer sampling points to approach the theoretical information upper bound, which also accounts for why its RMSE results in the Monte Carlo simulations more closely follow the PCRLB curves shown in Fig. \ref{fig:PCRLB}a.

\subsection{The experiment results}

\begin{figure}[ht]
    \centering
    \includegraphics[width=1\linewidth]{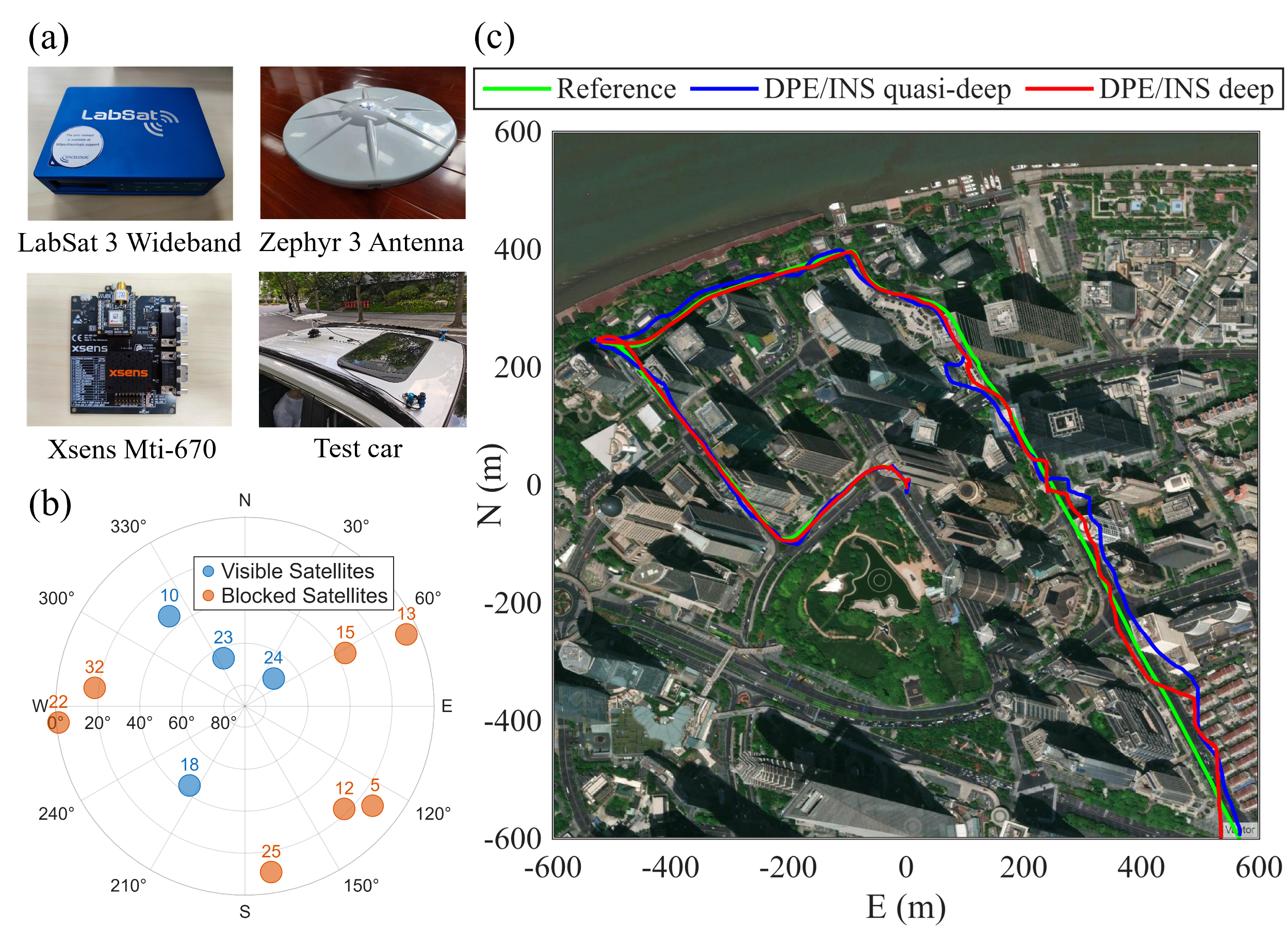}
    \caption{The experiment equipment, satellite sky plot, and DPE/INS deep integration results for the vehicle experiment}
    \label{fig:Test_setup_trajectory}
\end{figure}

The vehicle experiment is conducted on September 3, 2022, in Lujiazui CBD in Shanghai, China. The route trajectory is a typical urban canyon scenario which is appropriate to verify the effectiveness of the proposed DPE/INS deep integration framework. The test lasted for over 400 s. A LabSat 3 Wideband \cite{ref:Labsat} and a Trimble Zephyr 3 Antenna are used to collect the IF data at 30.69 MHz sampling frequency and 2-bit quantization. The IMU data is captured by the Xsens Mti-670 board at a sampling rate of 400Hz. The gyroscope bias stability and noise density are ${{8^\circ }/{\rm{h}}}$ and $0.007^\circ/s/ {\sqrt {Hz} }$, while $0.015{\rm{mg}}$ and ${{0.06{\rm{mg}}} / {\sqrt {\rm{Hz}} }}$ for accelerometers bias stability and noise density \cite{ref:Xsens}. For performance validation, the ground reference is generated using a commercial multi-sensor fusion system supported by the Xsens Mti-670 board, which provides the nominal position accuracy of 1 m and velocity accuracy of 0.05 m/s. The above equipment and the test vehicle are shown in Fig. \ref{fig:Test_setup_trajectory}a. 

As shown in Fig. \ref{fig:Test_setup_trajectory}b, there are a total of 11 GPS satellites. For a fair comparison, we selected four satellites, PRN 10, PRN 18, PRN 23, PRN 24, that remain observable throughout the entire experiment to obtain the DPE/INS deep integration results, as well as the quasi-deep integration results from our previous work for comparison, as presented in Fig. \ref{fig:Test_setup_trajectory}c. It can be observed that the deep integration results are closer to the reference trajectory most of the time, and the overall trajectory is also smoother.

\begin{figure}[ht]
    \centering
    \includegraphics[width=1\linewidth]{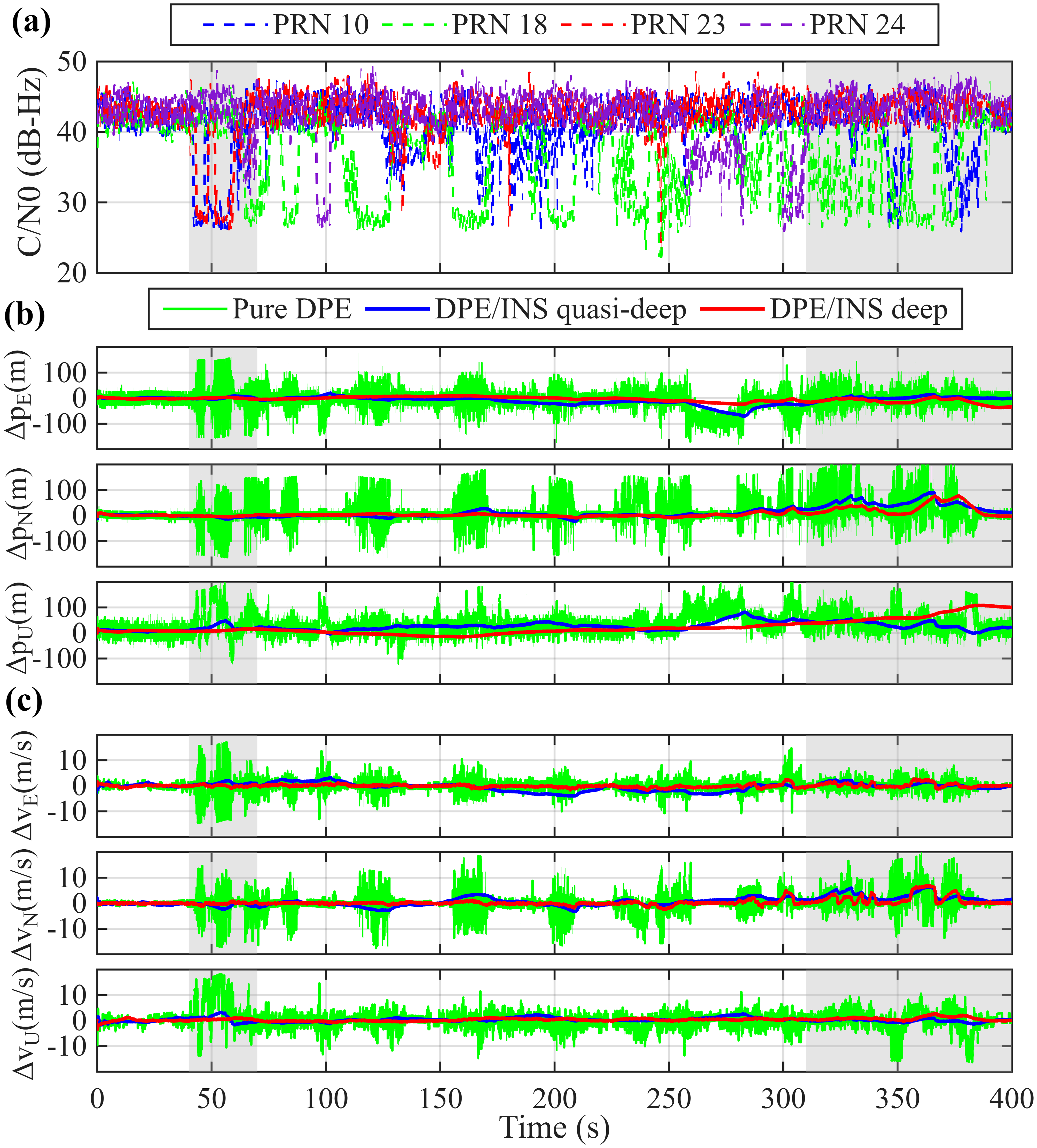}
    \caption{$C/N_0$ of each satellite and the position and velocity errors of the DPE/INS deep integration for the vehicle experiment}
    \label{fig:CN0_error}
\end{figure}

The $C/N_0$ of the selected satellites and the position and velocity errors along each axis during the vehicle experiment are further given in Fig. \ref{fig:CN0_error}. During the challenging segment around 40-70s, the quasi-deep integration suppresses part of pure-DPE fluctuation, reducing the position RMSE from 119.18 m to 26.37 m. In contrast, the proposed deep integration further reduces it to 11.46 m. A similar trend is observed in velocity estimation, where the 3D velocity RMSE is reduced from 2.14 m/s for the quasi-deep integration to 0.78 m/s for the proposed deep integration. These results indicate that embedding the INS prior directly into the DPE optimization is effective for suppressing short-term fluctuations under challenging GNSS conditions.

In contrast, the final stage from 310 s to 400 s shows a different error pattern. In this interval, the pure DPE error fluctuation exhibits a longer overall duration and displays a pronounced positive bias characteristic, suggesting that the correlation measurements are affected by persistent non-line-of-sight (NLOS) or multipath-induced systematic errors. Under this condition, both the quasi-deep and deep integration schemes gradually drifted under the influence of corrupted DPE information. This comparison indicates that the proposed deep integration improves robustness against short-term disturbances, but it cannot fully reject long-term and biased correlation measurements. Further improvement therefore requires explicitly modeling the influence of NLOS/multipath on the correlation values, so that biased correlation measurements can be detected or down-weighted in the optimization.

\begin{figure}[ht]
    \centering
    \includegraphics[width=1\linewidth]{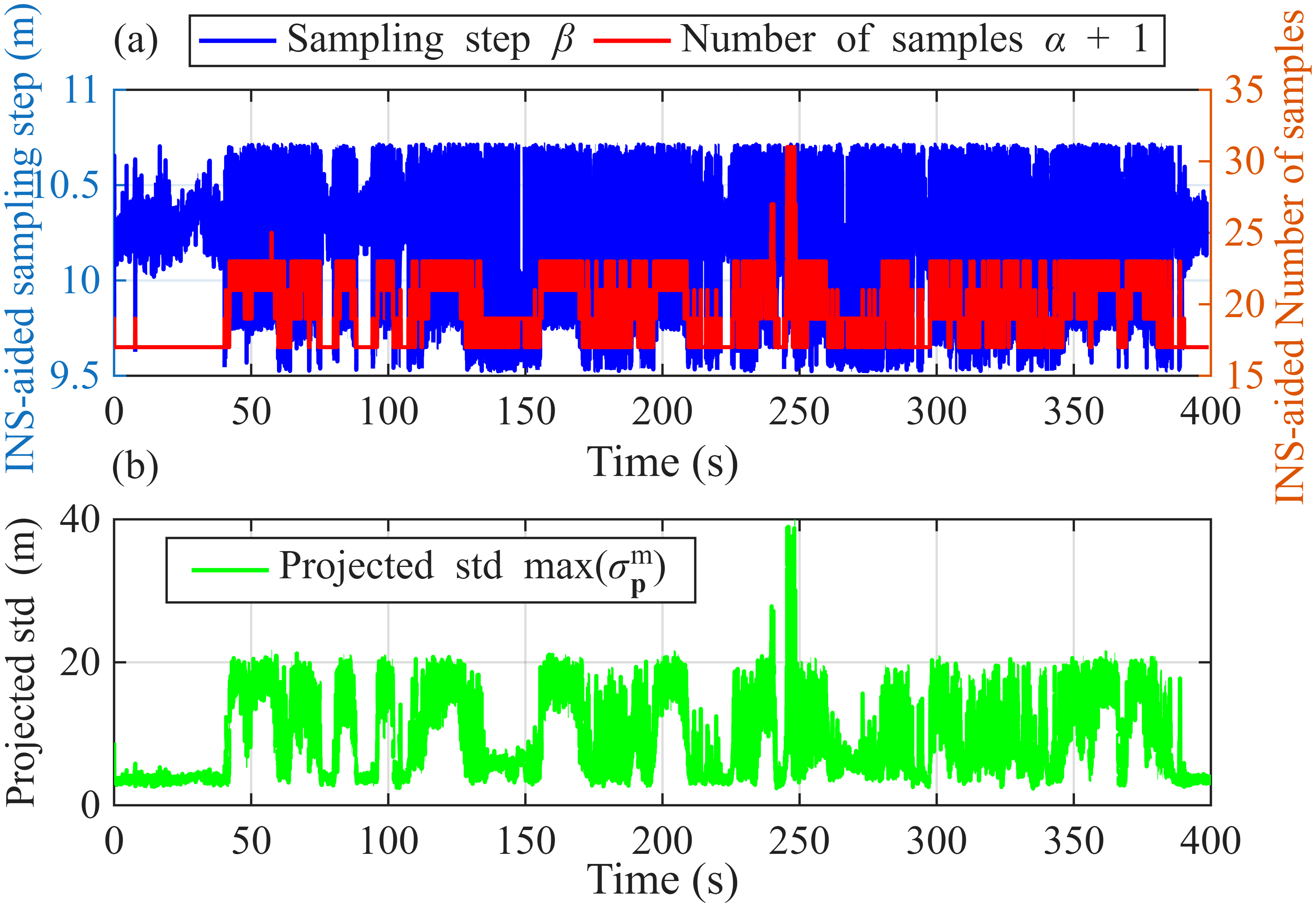}
    \caption{The selection of sampling parameters with INS aiding during the vehicle experiment}
    \label{fig:Los_sampling}
\end{figure}

Fig. \ref{fig:Los_sampling} presents the adaptive sampling behavior in the proposed deep integration framework. With INS aiding, the sampling parameters are adjusted according to the total projected uncertainty along the LOS direction, which combines the DPE prior uncertainty and the INS prediction uncertainty. In the experiment, the projected standard deviation along the LOS direction varies from 2.34 m to 39.85 m. Correspondingly, the number of sampling points is adaptively adjusted between 17 and 31, while the sampling step remains within a stable range of approximately 9.52 m-10.71 m. This strategy allows the sampling region to expand when the projected uncertainty increases, improving robustness in degraded environments, while avoiding unnecessary dense sampling when the uncertainty is small. Therefore, the adaptive sampling mechanism helps maintain both stability and computational efficiency for the DPE/INS deep integration.

\section{CONCLUSION}

 {This paper extends our previous navigation-domain DPE/INS quasi-deep integration to a correlation-domain DPE/INS deep integration framework based on joint optimization. The framework jointly estimates the navigation state and updates its covariance using LOS-sampled correlation measurements and INS prior information. The PCRLB analysis relates a specified fusion accuracy to the required correlation information and guides sampling design together with INS uncertainty coverage constraint.} Urban experiments demonstrate that, under equivalent conditions, the proposed method reduces position and velocity errors over 50\% in challenging scenarios compared with DPE/INS quasi-deep integration in navigation domain. Future work will extend this framework to incorporate additional sensors such as camera and LiDAR, fully exploiting the potential of the GNSS correlation domain in multi-source integrated navigation.

\begin{IEEEbiography}[{\includegraphics[width=1in,height=1.25in,clip,keepaspectratio]{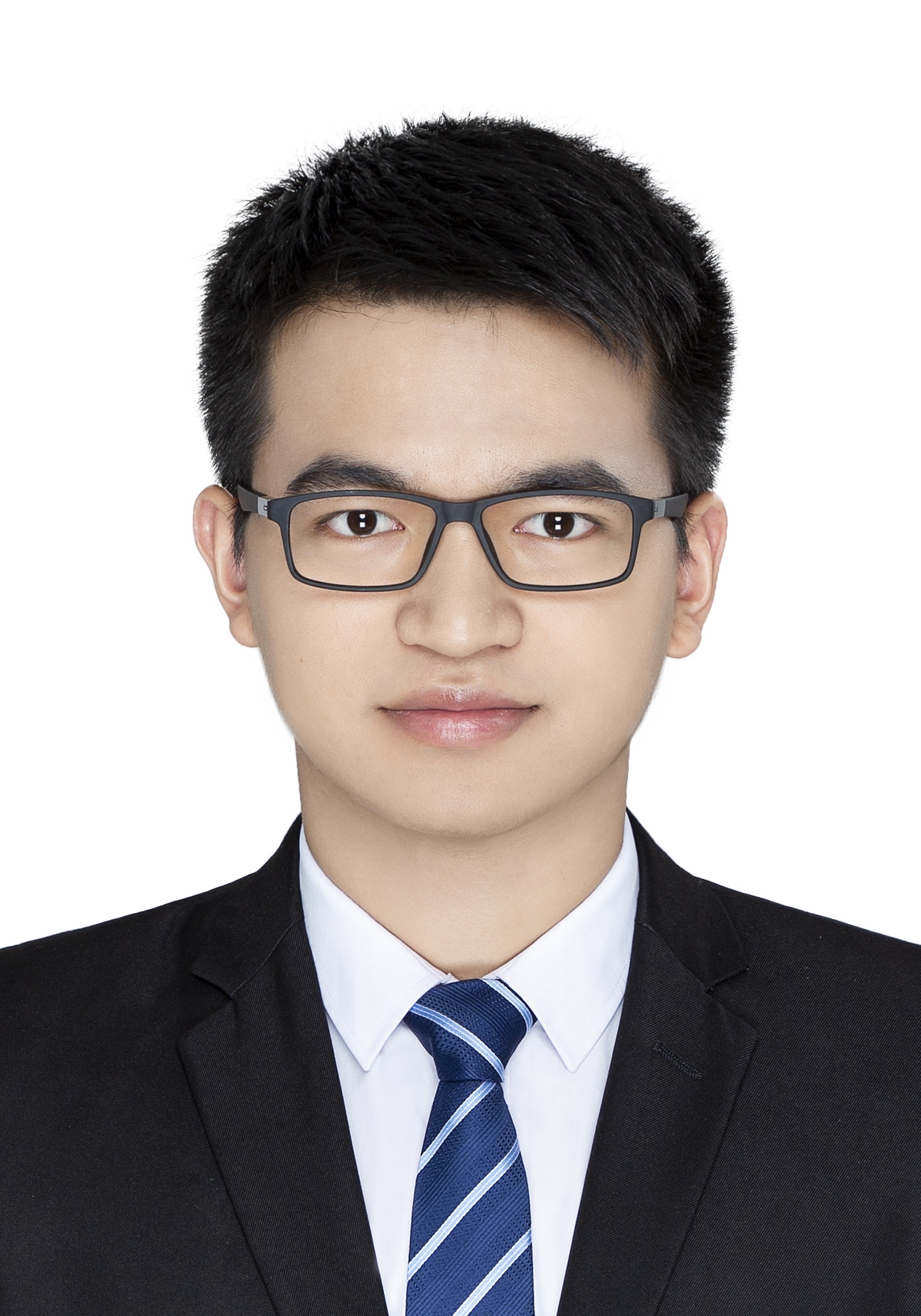}}]{Wei Gao} received his B.E. and M.E. degrees from Beihang University, China, in 2019 and 2022, respectively. He is currently a Ph.D. candidate at the School of Aeronautics and Astronautics, Shanghai Jiao Tong University (SJTU). His research interests focus on GNSS/INS integrated navigation and GNSS signal processing.
\end{IEEEbiography}

\begin{IEEEbiography}[{\includegraphics[width=1in,height=1.25in,clip,keepaspectratio]{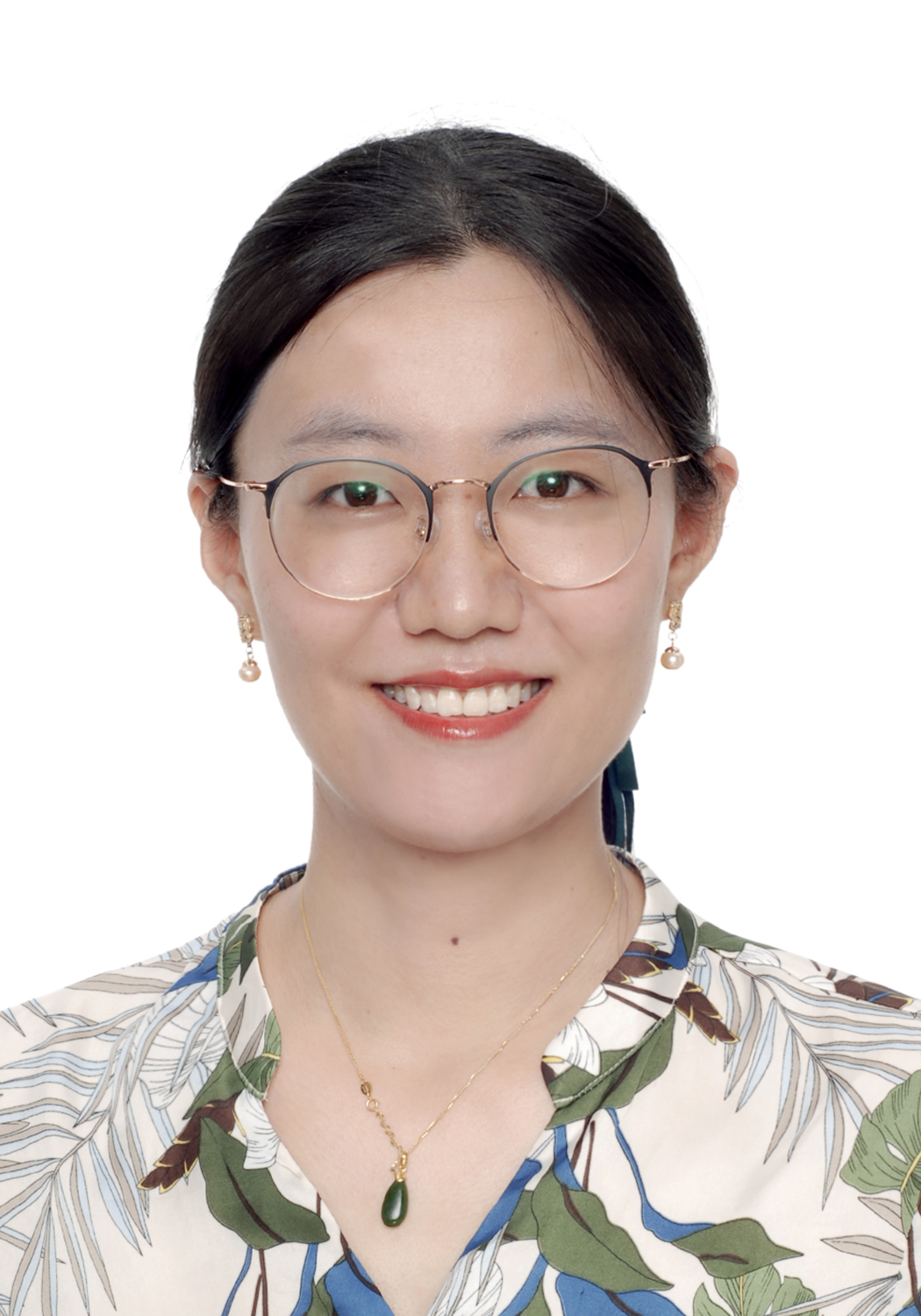}}]{Rong Yang} (Senior Member, IEEE) is an associate professor in the School of Aeronautics and Astronautics, Shanghai Jiao Tong University, China. She received her Ph.D. at Nanyang Technological University and was a Research Associate at the University of Colorado, Boulder. Her research focuses on GNSS signal processing for navigation in challenging environments. She serve as an associate editor for IEEE TAES. She received ION Per Enge early achievement award and IEEE AESS Early Career Award in 2025.

\end{IEEEbiography}

\begin{IEEEbiography}[{\includegraphics[width=1in,height=1.25in,clip,keepaspectratio]{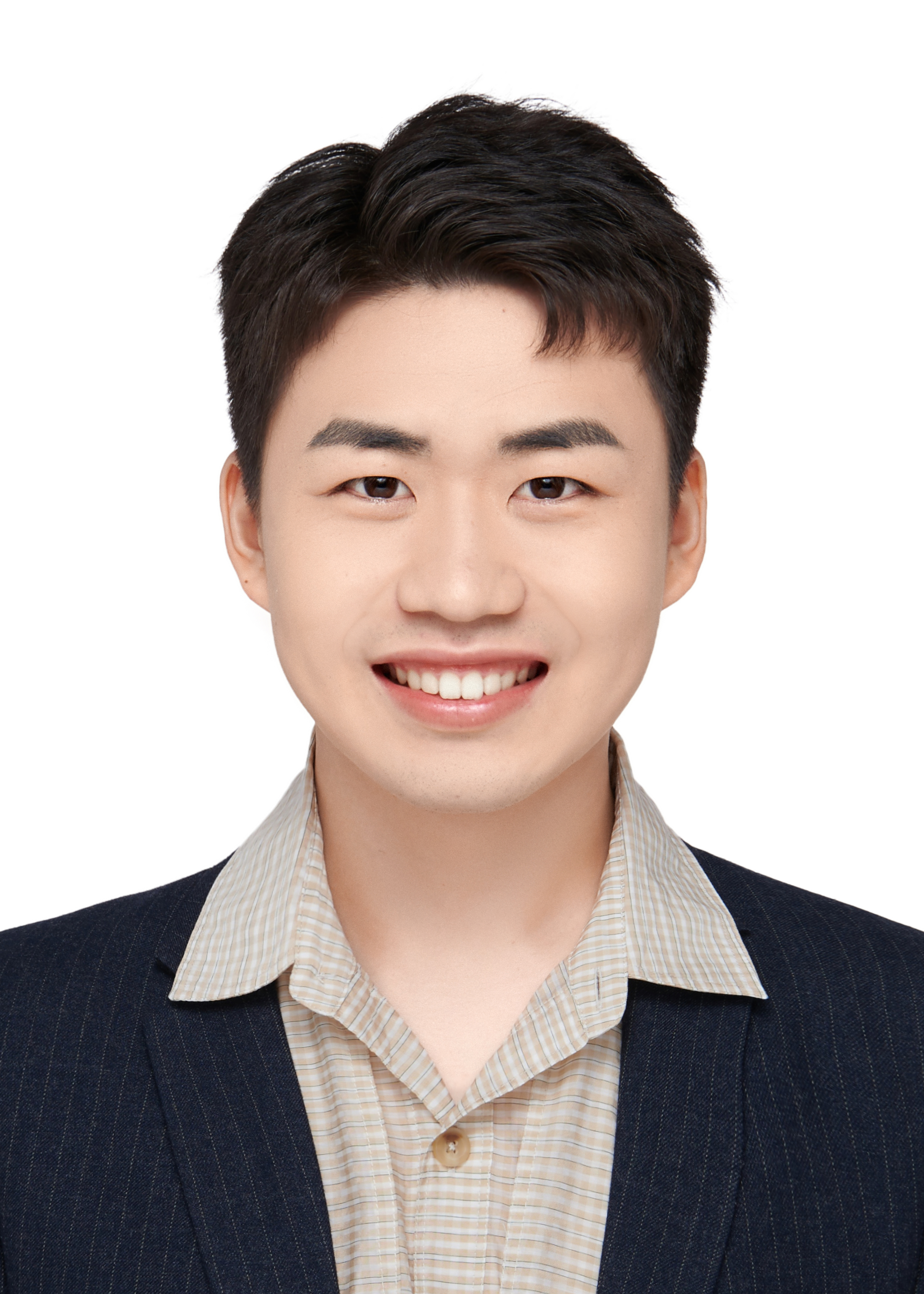}}] {Jihong Huang} received his Ph.D. in the School of Aeronautics and Astronautics, Shanghai Jiao Tong University (SJTU) in 2024. His research focuses on the application of GNSS reflectometry, multipath detection and mitigation, and direct position estimation (DPE).
\end{IEEEbiography}

\begin{IEEEbiography}[{\includegraphics[width=1in,height=1.25in,clip,keepaspectratio]{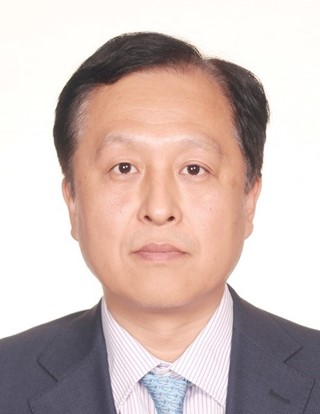}}]{Xingqun Zhan} (Senior Member, IEEE) received his Ph.D. from Harbin Institute of Technology (HIT), China, in 1999 and is currently a professor of the School of Aeronautics and Astronautics at Shanghai Jiao Tong University (SJTU). His research interests focus on navigation integrity, multi-sensor integration, and cooperative navigation.
\end{IEEEbiography}
\end{document}